\documentclass{emulateapj}
\usepackage{epsfig,natbib}
\usepackage{graphicx}
\received{2012 December 16} 
\accepted{2013 February 8} 
\slugcomment{Accepted by ApJ}
\newcommand{\wspitzer}{\emph{Warm-Spitzer}}
\newcommand{\spitzer}{\emph{Spitzer}}
\newcommand{\kms}{\ensuremath{\rm km\,s^{-1}}}
\newcommand{\ms}{\ensuremath{\rm m\,s^{-1}}\ }
\newcommand{\mse}{\ensuremath{\rm m\,s^{-1}}}
\newcommand{\gcmc}{\ensuremath{\rm g\,cm^{-3}}}
\newcommand{\teff}{\ensuremath{T_{\rm eff}}}
\newcommand{\logg}{\ensuremath{\log{g}}}
\newcommand{\vsini}{\ensuremath{v \sin{i}}}
\newcommand{\feh}{ \rm [Fe/H]}

\newcommand{\rsun}{\ensuremath{R_\sun}}
\newcommand{\msun}{\ensuremath{M_\sun}}
\newcommand{\lsun}{\ensuremath{L_\sun}}

\newcommand{\rstar}{\ensuremath{R_\star}}
\newcommand{\mstar}{\ensuremath{M_\star}}
\newcommand{\rprs}{\ensuremath{R_{\rm P}/R_\star}}
\newcommand{\esinw}{\ensuremath{e \sin (w)}}
\newcommand{\ecosw}{\ensuremath{e \cos (w)}}
\newcommand{\loggstar}{\ensuremath{\logg_\star}}
\newcommand{\lstar}{\ensuremath{L_\star}}

\newcommand{\rhostar}{\ensuremath{\rho_\star}}

\newcommand{\rpl}{\ensuremath{R_{\rm P}}}
\newcommand{\mpl}{\ensuremath{M_{\rm P}}}

\newcommand{\rhopl}{\ensuremath{\rho_{\rm P}}}
\newcommand{\loggpl}{\ensuremath{\logg_{\rm P}}}
\newcommand{\teq}{\ensuremath{T_{\rm eq}}}

\newcommand{\rearth}{\ensuremath{R_{\earth}}}
\newcommand{\mearth}{\ensuremath{M_{\earth}}}

\newcommand{\msini}{\ensuremath{M_{\rm P} \sin{i}}}

\newcommand{\koib}{Kepler-68b}
\newcommand{\koic}{Kepler-68c}
\newcommand{\koid}{Kepler-68d}
\newcommand{\starname}{Kepler-68}
\newcommand{\planetb}{Kepler-68b}
\newcommand{\planetc}{Kepler-68c}
\newcommand{\planetd}{Kepler-68d}
\newcommand{\kicid}{KIC~11295426}
\newcommand{\tmid}{2MASS J19240775+4902249}

\def\CM{{\cal M}}
\def\CP{{\cal P}}
\def\muHz{\,\mu{\rm Hz}}
\newcommand{\kicra}{\ensuremath{19^{\mathrm{h}}24^{\mathrm{m}}07^{\mathrm{s}}.75}}
\newcommand{\kicdec}{\ensuremath{+49^{\circ}02'25''.0}}
\newcommand{\kepmag}{10.00}

\newcommand{\teffSMEOrig}{\ensuremath{5754 \pm 44}}    % HIRES
\newcommand{\fehSMEOrig}{\ensuremath{+0.10 \pm 0.04}}   % HIRES
\newcommand{\loggSMEOrig}{\ensuremath{4.2 \pm 0.1}}   % HIRES
\newcommand{\vsiniSMEOrig}{\ensuremath{0.5 \pm 0.5}}    % HIRES

\newcommand{\teffSME}{\ensuremath{5793 \pm 44}}     % HIRES

\newcommand{\fehSME}{\ensuremath{+0.12 \pm 0.04}}   % HIRES
\newcommand{\loggSME}{\ensuremath{4.281 \pm 0.06}}   % HIRES
\newcommand{\vsiniSME}{\ensuremath{0.5 \pm 0.5}}     % HIRES
\newcommand{\teffsys}{\ensuremath{5793 \pm 74}}     % HIRES
\newcommand{\fehsys}{\ensuremath{+0.12 \pm 0.074}}   % HIRES
\newcommand{\mstarKASC}{\ensuremath{1.079 \pm 0.051}}
\newcommand{\rstarKASC}{\ensuremath{1.243 \pm 0.019}}
\newcommand{\rhostarKASC}{\ensuremath{0.7903 \pm 0.0054}}
\newcommand{\loggKASC}{\ensuremath{4.281 \pm 0.008}}

\newcommand{\lumKASC}{\ensuremath{1.564 \pm 0.141}}
\newcommand{\absMag}{\ensuremath{4.34 \pm 0.09}}
\newcommand{\ageKASC}{\ensuremath{6.3 \pm 1.7}}
\newcommand{\distance}{\ensuremath{135 \pm 10}}

\newcommand{\periodb}{\ensuremath{5.398763 \pm 0.000004}}
\newcommand{\epochb}{\ensuremath{2455006.85729 \pm 0.00042}}
\newcommand{\depthb}{\ensuremath{345.6 \pm 1.5}}
\newcommand{\durationb}{\ensuremath{3.459 \pm 0.009}}
\newcommand{\impactb}{\ensuremath{0.45 \pm 0.17}}

\newcommand{\scaledSemiMajb}{\ensuremath{10.68 \pm 0.14}}
\newcommand{\scaledPlanetRadiusb}{\ensuremath{0.01700 \pm 0.00046}}
\newcommand{\inclinationb}{\ensuremath{87.60 \pm 0.90}}

\newcommand{\periodc}{\ensuremath{9.605085 \pm 0.000072}}
\newcommand{\epochc}{\ensuremath{2454969.3805 \pm 0.0041}}
\newcommand{\depthc}{\ensuremath{53.1 \pm 2.3}}
\newcommand{\durationc}{\ensuremath{3.09 \pm 0.09}}

\newcommand{\scaledSemiMajc}{\ensuremath{15.68 \pm 0.20}}
\newcommand{\scaledPlanetRadiusc}{\ensuremath{0.00703 \pm 0.00025}}
\newcommand{\impactc}{\ensuremath{0.84 \pm 0.11}}
\newcommand{\inclinationc}{\ensuremath{86.93 \pm 0.41}}

\newcommand{\semiAmpb}{\ensuremath{2.63 \pm 0.71}}
\newcommand{\eccb}{\ensuremath{0}}
\newcommand{\gammaVelb}{\ensuremath{-20.9 \pm 0.1}}

\newcommand{\semiAmpc}{\ensuremath{1.25^{+0.65}_{-0.95}}}

\newcommand{\periodd}{\ensuremath{580\pm 15}}
\newcommand{\semiAmpd}{\ensuremath{19.9 \pm 0.75}}
\newcommand{\eccd}{\ensuremath{0.18 \pm 0.05}}

\newcommand{\mplanetb}{\ensuremath{8.3^{+2.2}_{-2.4}}}
\newcommand{\rplanetb}{\ensuremath{2.31^{+0.06}_{-0.09}}}
\newcommand{\rhoplanetb}{\ensuremath{3.32^{+0.86}_{-0.98}}}
\newcommand{\loggplanetb}{\ensuremath{3.14 \pm 0.11}}
\newcommand{\semiMajb}{\ensuremath{0.06170 \pm 0.00056}}

\newcommand{\teqb}{\ensuremath{1280 \pm 90}}

\newcommand{\mplanetc}{\ensuremath{4.8^{+2.5}_{-3.6}}}
\newcommand{\rplanetc}{\ensuremath{0.953^{+0.037}_{-0.042}}}
\newcommand{\rhoplanetc}{\ensuremath{28.^{+13.}_{-23.}}}

\newcommand{\semiMajc}{\ensuremath{0.09059 \pm 0.00082}}

\newcommand{\mplanetd}{\ensuremath{0.947 \pm 0.035}}

\newcommand{\semiMajd}{\ensuremath{1.40 \pm 0.03}}

\newcommand{\ek}{\emph{Kepler}}
\newcommand{\blender}{{\tt BLENDER}}

\shortauthors{Gilliland {\it et~al.\/}}
\shorttitle{Kepler-68: Three Planets}

\begin{document}
\pagenumbering{arabic}

\title{Kepler-68: Three Planets, One With a Density Between that of Earth and Ice Giants}

\author{Ronald~L.~Gilliland\altaffilmark{1}, % Penn State, gillil@stsci.edu
Geoffrey~W.~Marcy\altaffilmark{2},  % Berkeley, gmarcy@berkeley.edu
Jason~F.~Rowe\altaffilmark{3}, % Ames, jasonfrowe@gmail.com
Leslie~Rogers\altaffilmark{4}, % Caltech, larogers@caltech.edu
Guillermo~Torres\altaffilmark{5}, % Harvard, gtorres@cfa.harvard.edu
Francois~Fressin\altaffilmark{5}, % Harvard, ffressin@cfa.harvard.edu
Eric~D.~Lopez\altaffilmark{6}, % UCSC, edlopez@ucsc.edu
Lars~A.~Buchhave\altaffilmark{7}, % Copenhagen, buchhave@astro.ku.dk
J{\o}rgen~Christensen-Dalsgaard\altaffilmark{8,9}, % Aarhus and HAO, jcd@phys.au.dk
Jean-Michel~D\'esert\altaffilmark{4}, % caltech, desert@caltech.edu
Howard~Isaacson\altaffilmark{2},  % Berkeley, hisaacson@berkeley.edu
Jon~M.~Jenkins\altaffilmark{10}, % SETI Inst, jon.m.jenkins@nasa.gov
Jack~J.~Lissauer\altaffilmark{3}, %Ames, Jack.J.Lissauer@nasa.gov
William~J.~Chaplin\altaffilmark{11}, % Birmingham wjc@bison.ph.bham.ac.uk
Sarbani~Basu\altaffilmark{12}, % Yale  sarbani.basu@yale.edu
Travis~S.~Metcalfe\altaffilmark{13}, % Boulder  travis@spacescience.org
Yvonne~Elsworth\altaffilmark{11}, % Birmingham  ype@bison.ph.bham.ac.uk
Rasmus~Handberg\altaffilmark{8}, % Aarhus  rasmush@phys.au.dk
Saskia~Hekker\altaffilmark{14}, % Amsterdam  s.hekker@uva.nl
Daniel~Huber\altaffilmark{3}, % Ames, daniel.huber@nasa.gov
Christoffer~Karoff\altaffilmark{8}, % Aarhus  karoff@phys.au.dk
Hans~Kjeldsen\altaffilmark{8}, % Aarhus, hans@phys.au.dk
Mikkel~N.~Lund\altaffilmark{8}, % Aarhus mikkelnl@phys.au.dk
Mia~Lundkvist\altaffilmark{8}, % Aarhus lundkvist@phys.au.dk
Andrea~Miglio\altaffilmark{11}, % Birmingham miglioa@bison.ph.bham.ac.uk
David~Charbonneau\altaffilmark{5}, % Harvard, dcharbonneau@cfa.harvard.edu
Eric~B.~Ford\altaffilmark{15}, % Florida, eford@astro.ufl.edu
Jonathan~J.~Fortney\altaffilmark{6}, % UCSC, jfortney@ucolick.org
Michael~R.~Haas\altaffilmark{3}, % Ames, michael.r.haas@nasa.gov
Andrew~W.~Howard\altaffilmark{16},  % Hawaii, howard@ifa.hawaii.edu
Steve~B.~Howell\altaffilmark{3}, % NOAO, steve.b.howell@nasa.gov
Darin~Ragozzine\altaffilmark{15}, % Harvard, darin.ragozzine@gmail.com
Susan~E.~Thompson\altaffilmark{10}, % SETI  susan.e.thompson@nasa.gov
}
\altaffiltext{1}{Department of Astronomy, and Center for Exoplanets and Habitable Worlds,
The Pennsylvania State University, 525 Davey Lab, University Park, PA 16802; gillil@stsci.edu}
\altaffiltext{2}{University of California, Berkeley, CA 94720}
\altaffiltext{3}{NASA Ames Research Center, Moffett Field, CA 94035}
\altaffiltext{4}{California Institute of Technology, Pasadena, CA  91125}
\altaffiltext{5}{Harvard-Smithsonian Center for Astrophysics, 60 Garden Street, Cambridge, MA 02138}
\altaffiltext{6}{University of California, Santa Cruz, CA 95064}
\altaffiltext{7}{Niels Bohr Institute, Copenhagen University, Denmark}
\altaffiltext{8}{Stellar Astrophysics Centre, Dept. of Physics and Astronomy
DK-8000  Aarhus C, Denmark}
\altaffiltext{9}{High Altitude Observatory, National Center for Atmospheric Research, Boulder, CO 80307}
\altaffiltext{10}{SETI Institute/NASA Ames Research Center, Moffett Field, CA 94035}
\altaffiltext{11}{School of Physics and Astronomy, University of Birmingham, Edgbaston, Birmingham B15 2TT, UK}
\altaffiltext{12}{Yale University, 260 Whitney Ave., New Haven, CT 06511}
\altaffiltext{13}{White Dwarf Research Corporation, Boulder, CO 80301}
\altaffiltext{14}{Astronomical Institute Anton Pannekoek, University of Amsterdam, 
1098 XH Amsterdam, Science Park 904, The Netherlands}
\altaffiltext{15}{University of Florida, Gainesville, FL 32611}
\altaffiltext{16}{Institute for Astronomy, University of Hawaii,
2680 Woodlawn Drive, Honolulu, HI  96822}

\begin{abstract}
NASA's {\em Kepler Mission} has revealed two transiting planets orbiting \starname.
Follow-up Doppler measurements have established the mass of the innermost
planet and revealed a third jovian-mass planet orbiting beyond
the two transiting planets.   \planetb, in a 5.4 day orbit, has $\mpl=\mplanetb$ \mearth,
$\rpl=\rplanetb$ \rearth, and $\rhopl=\rhoplanetb$ g cm$^{-3}$,
giving \planetb\ a density intermediate between that of
the ice giants and Earth.
\planetc \ is Earth-sized with a radius, $\rpl=\rplanetc$ \rearth\ and
transits on a 9.6 day orbit; validation of \koic\ posed unique challenges.
\planetd \ has an orbital
period of \periodd \ days and minimum mass of \msini = \mplanetd $M_J$.
Power spectra of the \ek \ photometry at 1-minute cadence
exhibit a rich and strong set of asteroseismic pulsation modes
enabling detailed analysis of the stellar interior.  Spectroscopy of
the star coupled with asteroseismic modeling of the multiple
pulsation modes yield precise measurements of stellar properties,
notably $\teff=\teffsys$ K, $\mstar=\mstarKASC$ \msun, $\rstar=\rstarKASC$ \rsun,
and $\rhostar = \rhostarKASC$ g cm$^{-3}$, all measured with fractional
uncertainties of only a few percent. Models of \planetb \ suggest it
is likely composed of rock and water, or has a H and He envelope
to yield its density $\sim$3 g cm$^{-3}$.
\end{abstract}

\keywords{planetary systems --- stars: fundamental parameters ---
stars: individual (\starname, \kicid, \tmid)}

%=====================================================================

\section{Introduction}

The NASA {\em Kepler Mission} has presented a catalog of over 2300 stars with
planet-like transit signatures \citep{boru11, bata12}.
Here we report a detailed study of \starname, a G-type main sequence star
harboring a transiting planet, \koib \ having a radius of
$\sim$2.5 \rearth\ and orbital period, $\sim$5.40 d.
We describe the detection of a second transiting planet that is
close in size to the Earth.  We carry out multiple follow-up
measurements of the star \starname\ (KIC 11295426), including additional \ek\ photometry,
ground-based spectroscopy and high resolution imaging,
{\em Spitzer Space Telescope} photometry, and Doppler measurements.
At Kepler magnitude, $Kp \, =\kepmag$, the star has
high enough flux for asteroseismic analysis of its stellar properties
using short cadence (see \citet{gill10}) \ek\ photometry, offering correspondingly
accurate measures of the stellar density, mass, and radius.   

The photometry of \starname\ and subsequent transit detections
of \koib \ and \koic \ are described in Section~\ref{sec:photometry}.
We present tests performed on the \ek\ photometry and images to rule out false
positives in Section~\ref{sec:dv}.  We present the follow-up observations,
including spectroscopy, high resolution imaging, {\em Spitzer Space Telescope} photometry,
and precision Doppler measurements,
leading to the support of \koib\ and \koic\ as planets in Section~\ref{sec:fop}.
We refer to the two transiting planets as \koib\ and \koic\
and the subsequent Doppler-detected outer planet as \koid.
We report a corresponding Doppler signal for \koib, but the radial velocity (RV)
measurements provide only an upper limit to the mass of \koic,
that is physically uninteresting.  We describe investigation of
false-positive scenarios with a \blender\ analysis \citep{torr11}
as described in Section~\ref{sec:blender}.  

The spectroscopy and asteroseismology yield stellar
properties discussed in Section~\ref{sec:star}.  The stellar density, mass,
and radius permit a detailed analysis of the light curve and
Doppler measurements, to give planet parameters, described in Section~\ref{sec:planet}.  
We also discuss the properties of \planetb\ from the standpoint of
theoretical models in Section~\ref{sec:planet}, especially regarding the planet's composition.
Its placement in a mass-radius diagram suggests a composition of
large amounts of rock and significant amounts of volatiles to yield the
observed density of 3 g cm$^{-3}$.  

With the changing status of stars during the course of the {\em Kepler Mission},
as planet candidates are discovered and confirmed as planets, the 
nomenclature used and recognized by diverse analysis groups evolves.
The star studied here is located at $\alpha$ = \kicra, $\delta$ = \kicdec~
and in the Kepler Input Catalog (KIC) was designated KIC 11295426.
The previously existing 2MASS ID was J19240775+4902249.
The Kepler Object of Interest (KOI) name was given when it appeared in
the \citet{boru11} exoplanet candidate list as KOI00246, or the more
commonly appearing KOI-246 as used herein.
%With the identification of clear stellar oscillations, the star
%became ``Harry" within the asteroseismology group per the 
%custom of adopting pet cat names.
KOI numbers were assigned per convention as: KOI-246.01 with
initial detection of the 5.4 day candidate transits, KOI-246.20
for the candidate detected based on non-\ek\ input, and then
KOI-246.02 for the second transiting planet candidate at 9.6 days.
With validation and confirmation of planets the star was given its final moniker,
Kepler-68, and the planets Kepler-68b,c,d -- from KOIs 246.01, 246.02
and 246.20 respectively.

\section{\ek\ Photometry}
\label{sec:photometry}

The \ek\ instrument is described in \citet{vanc09} and \citet{arga08}
while an overview of performance is presented in \citet{cald10b}
and \citet{jenk10b}.
Here we report the results from using 12 quarters of Kepler data.
The standard pipeline reduction of the photometry first yielded a
transit signal with a period of 5.40 d, consistent with a planet with
a size of approximately 2.5 \rearth \citep{boru11, bata12}.
Subsequent searches of the light curve alerted the \ek\ team to a
second transit signal with a period of $\sim$9.61 d and approximately Earth size.
Thus, the \ek\ photometry and pipeline reduction from \starname \ reveals
two periodic transit signals consistent with planets, hereafter called \koib \ and \koic .  

Independent transit searches of the \ek\ photometry have been carried
out by \citet{ofir12} and \citet{huang12} who also find evidence for \koib \ and \koic \ 
with the same period and transit depth within uncertainties.  For \koic \
\citet{ofir12} find a period of 9.60538 $\pm$ 0.00026 d and a planet radius of 0.86 \rearth.
They also find a single transit from a possible third planet with a
transit duration of $\sim$8 hr implying a period of 970 $\pm$ 50 days,
and planet size about 2.4 \rearth.  Thus there is a possibility that a
third planet transits \starname.
\ek\ observations of \starname\ 
are ongoing, including acquisition of short cadence data.
Inspection of data through Q13 has not shown further evidence 
for the several hundred day planet candidate.
At a period of nearly 1000 days the next transit would not be
expected until Q16 in early 2013.

Raw flux light curves for each quarter \citep{jenk10a} are
corrected for systematic errors, detrended, and stitched together to form
contiguous time series, and are then searched for transit signals
\citep{jenk10c}. We remove systematic errors, outliers and
intra-quarter discontinuities by co-trending against the photometry of
other stars in the vicinity of Kepler-68 using the Pre-Search Data
Conditioning (PDC) pipeline module as described in \citet{twic10b},
with updates as per \citet{stum12}; \citet{smit12}.

\begin{figure}
\begin{center}
\includegraphics[width=170mm]{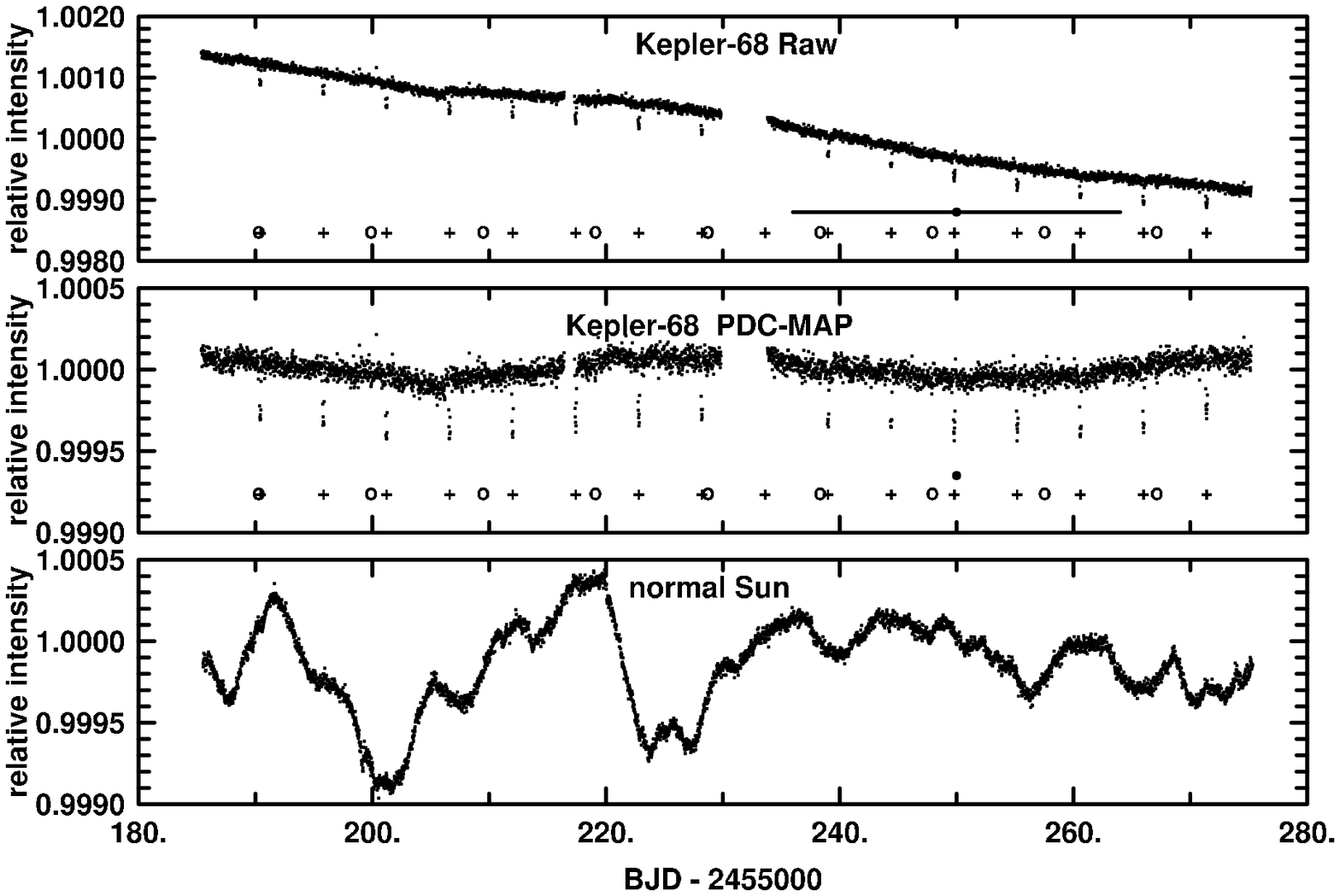}
\end{center}
\caption{Top panel shows the raw flux (SAP\_FLUX) time series for
\starname, after normalization by the median for the representative quarter Q4.
The plus signs flag central times of transits of \koib,
while the open circles flag transits of \koic.
The solid dot shows the expected position for a transit of the 
outer, RV-detected \koid\ with the horizontal bar showing the 1-$\sigma$
phase uncertainty -- no transit is seen.
PDC-MAP corrected flux time series produced by the \ek\ photometry
pipeline (PDCSAP\_FLUX) is shown in the middle panel.
The lower panel shows
a 90-day segment of SOHO VIRGO/SPM \citep{froh97} data from the green
channel scaled as discussed in \citet{gill11b} to match the \ek\ 
bandpass.  The solar data centered on 2005.52 were taken from 
a period of average variability.}
\label{fig:q4phot}
\end{figure}

Figure~\ref{fig:q4phot} shows the raw (SAP\_FLUX, the result
of simple aperture photometry) and corrected (PDCSAP\_FLUX) flux time series
for \starname\ during a representative quarter (Q4).
The largest remaining systematic errors in the PDC-MAP 
(\citet{stum12}; \citet{smit12}) processed data
are minor offsets following thermal transients after monthly 
pointing changes to telemeter data to the ground (very small for Q4).
The slow variation with a period of $\sim$50 days and amplitude 
of 0.0003 in this figure could either be intrinsic to the star,
or associated with imperfect removal of long-term drifts due to image 
motion (differential velocity aberration) that is present in the raw data.
It is clear that variations in \starname\ are smaller than
typical variations of the Sun, consistent with the slow rotation
and advanced age argued for in sections 6.1 and 6.2 respectively.
After filtering out transit events,
the measured relative standard deviation of the PDC-corrected,
long-cadence light curve is 21 ppm per 6 hour interval (CDPP --
the formal Combined Differential Photometric Precision
-- see \citet{jenk10b}, and \citet{chri12b}).
An expected instrument $+$ photon noise is computed for each flux in the timeseries.
The mean of the per (29.4-minute) cadence noise estimates
reported by the pipeline is 233 ppm.
Both raw (simple aperture sums) and corrected (PDC-MAP) light curves are available
at the Mikulski Archive for Space Telescopes (MAST)\footnote{\url{http://archive.stsci.edu/kepler}}
at the Space Telescope Science Institute.

\subsection{Transiting Planet Search}
\label{sec:tps}

The \planetb \ transits were identified by the Transiting Planet Search
(TPS) pipeline module that identifies periodic reductions in flux having a
duration of hours, each corresponding to a transit of a prospective planet.
The algorithm is a wavelet-based, adaptive matched filter \citep{jenk10c}.
TPS then identifies a time series of single ``events", each having
an associated ``single 
event statistic (SES)" that represents the probability that a transit is present.
The SES from each transit are combined into multiple event statistics
(MES) by folding them at trial orbital periods ranging from 0.5 days
to as long as half the data coverage interval.

\koib\ was identified by TPS in each quarter of data with a multiple
event statistic $>15\sigma$.

Multi-quarter searching for transits was used.  The transit depth, duration, period,
and epoch are derived from physical modeling (see Section~\ref{sec:planet})
using all of the available data.  \koib\ is characterized as a \depthb\ ppm
dimming lasting \durationb\ hours with transit ephemeris
of T\,[BJD]\,$=\,\epochb\,+\,N*\periodb$ days. 
The longer-period transits of \koic\ were identified by non-pipeline inspections.
\koic\ is characterized as a \depthc\ ppm dimming lasting \durationc\ hours
and an ephemeris T\,[BJD]\,$=\,\epochc\,+\,N*\periodc$ days. 

\section{Data Validation}
\label{sec:dv}

Signals that mimic planet transits are also found by TPS.
All ``threshold crossing events" identified by TPS are subjected
to assessment of standard vetting products that allow 
separate disposition of clear false positives before bestowing
the KOI moniker.
Most false positives can be identified by judicious assessment of the quality of
the transit-planet model fit to the photometry and by searching for
astrometric displacements of the photocenter between times out of
transit and in-transit.  True transiting planets should exhibit
photometry that is well fit by a transiting planet model and they
should show little, if any, astrometric displacement during transit
(depending on neighboring stars).   Such ``Data Validation" techniques
are described in \cite{bata10} and in \cite{bata11}.
Both \koib\ and \koic\ passed all such tests, conferring KOI planet candidate status for
both of them, as will be further discussed next.

\begin{figure}
\begin{center}
\includegraphics[height=110mm]{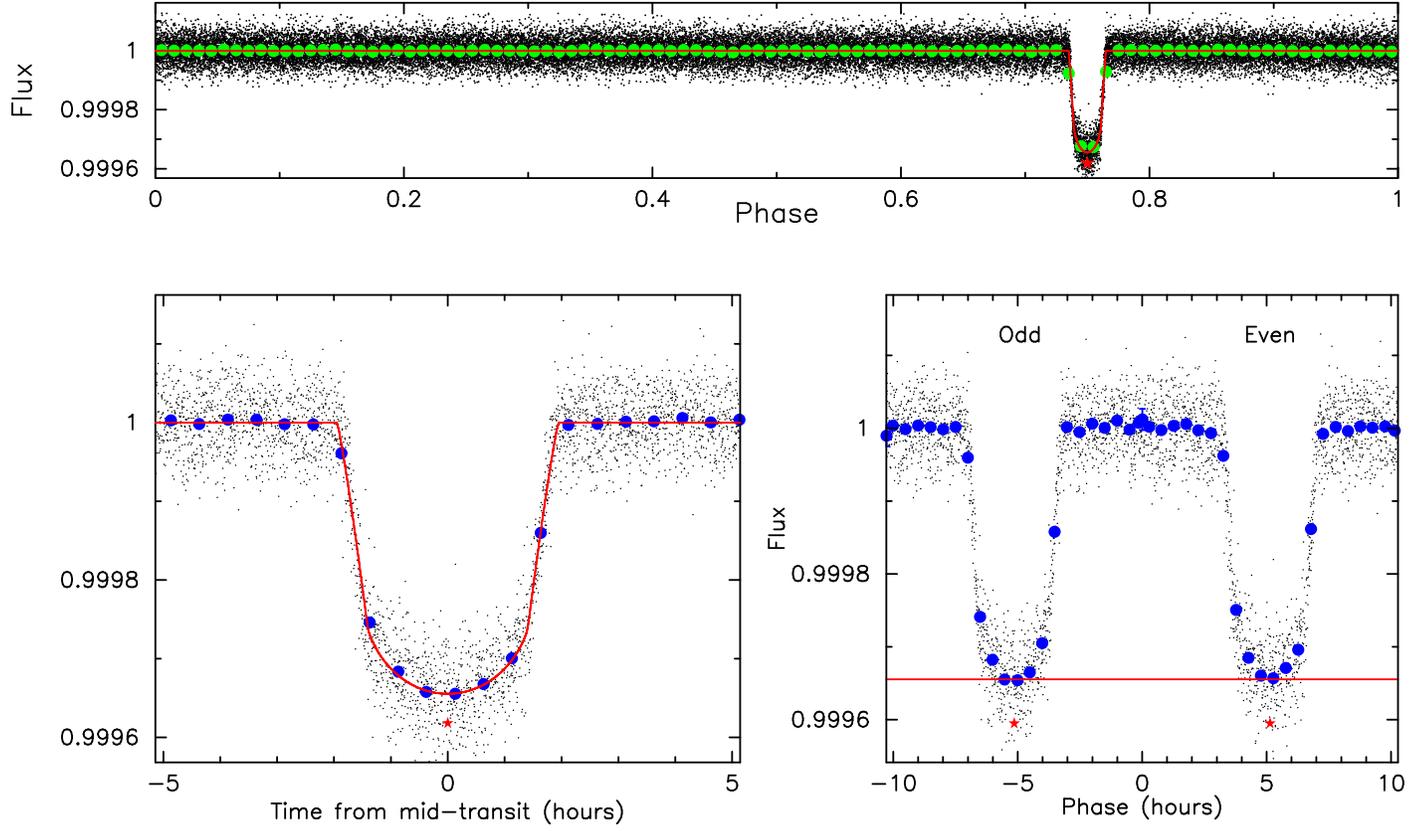}
\end{center}
\caption{The upper panel shows the de-trended time series
over Q1-Q11 (without Q3 as discussed in the text) after folding on the
5.39877 day period of \koib\ such that the transit falls at phase
0.75.  The signal for \koic\ has also been subtracted.
Green points show the data binned in 0.01 phase intervals,
the red line is the best fitting transit model.
The lower left panel provides detail on the phased and
folded light curve at the position of the transit; solid dots
indicate 30-minute averages and the solid line is the best fitting
transit light curve fit.  The lower right panel details Odd and 
Even numbered transits individually co-added.
Red stars mark center-of-transit times.}
\label{fig:vett01}
\end{figure}

Figure~\ref{fig:vett01} shows several representations of data relevant
to judging the \koib\ signal.
In some early quarters
the optimal aperture did not encompass all of the flux,
especially for saturated targets like \starname.
This leads to suppressed variation and is the reason we 
did not use the Q3 data.
The upper panel shows the phase folded
light curve for \koib, after de-trending and subtraction
of the \koic\ transit signal.
A common type of false positive is a background eclipsing
binary blended with the target star; in such cases a secondary eclipse
is often seen.  No evidence of a secondary eclipse is seen here, nor
turned up in automated searches.  The lower left panel shows
that the phase folded data for \koib\ are fit very well by our transit model.
Another signature of a possible false positive associated with an eclipsing
binary (either the target star itself as a grazing eclipse or as a 
blended background system) relies on searching for subtle differences of
depth between alternate transits -- a binary with twice the listed period
and primary and secondary eclipses of slightly unequal depths are easily
seen in data of this quality.  From all aspects of lightcurve inspection
\koib\ is entirely clean.

\begin{figure}
\begin{center}
\includegraphics[height=110mm]{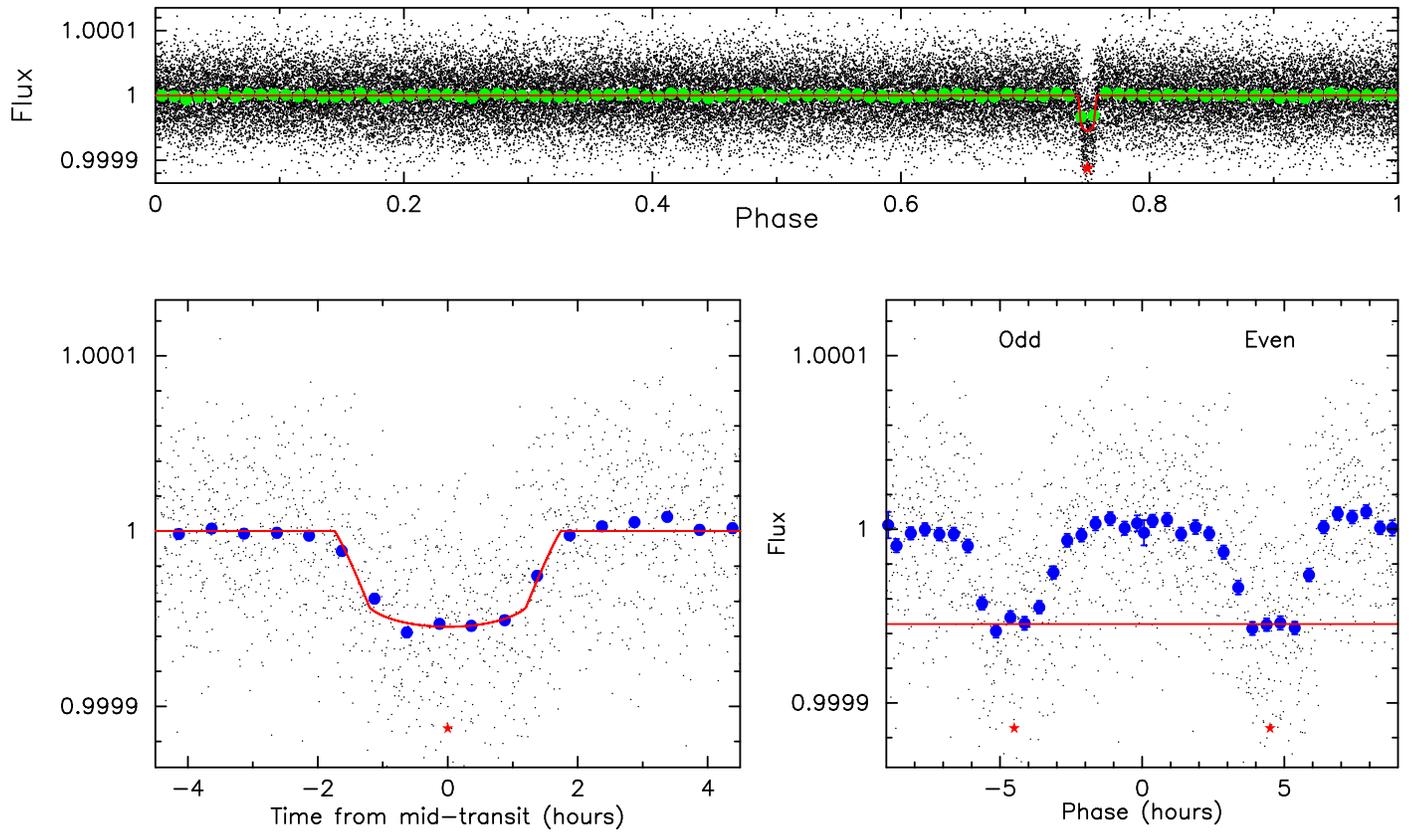}
\end{center}
\caption{Same as Figure 2, but for \koic\ after removal
of the \koib\ transits.}
\label{fig:vett02}
\end{figure}

Figure~\ref{fig:vett02} addresses \koic\ showing analogous vetting products for the
time series as Figure~\ref{fig:vett01}, but with the transit signal
of \koib\ subtracted out.  Again, there is no clear evidence for a
secondary eclipse, or depth differences for alternately averaged
transits, either of which would be suggestive of a false positive
interpretation.  The question of a secondary eclipse feature
will, however, be explored at depth in the \blender\ context
in Section 5.  This transit, that is only $\sim$60\% as deep as
a true Earth analog, obviously appears at high signal-to-noise 
in the phase folded data and is well fit by our planet transit model.

Centroid analyses based on assessing the in- and out-of-transit 
signal positions can be particularly powerful, (e.g. see \citet{bata11}
for application to Kepler-10b).  The difference of average images
taken out-of-transit, minus a similar average of images taken during
transit for an unsaturated target yields a PSF at the inherent 
source position (coincident within errors with the target star for 
a clean candidate).  Since \starname\ is saturated, the
centroid analyses are much less discerning with the inherent loss
of spatial information following saturation.  For a strongly saturated
target changes of flux are concentrated at the end of bleed columns, 
while the central pixels in the bleed
trail remain, well, saturated.

The KIC shows the location of a fainter
(magnitude not available in KIC, but derived below in Section 4.2.1
from AO imaging) star offset by about 11\arcsec~ in such a way that
it is almost precisely in the detector column direction from 
\starname.  In most quarters \starname\ is roughly centered on
a pixel in the column direction leading to bleeding that 
encompasses this secondary star, KIC 11295432, in which 
circumstance using the \ek\ data to discern the true source of
transits between the two stars is not possible.  In Q9
both \starname\ and the fainter neighbor KIC 11295432 are 
nearly centered between columns, and the bleeding terminates
before reaching the latter.
Table~\ref{tab:difftab} details pixel values in both the direct
out-of-transit image at the left, and for the difference images
of \koib\ and \koic\ in the central and right blocks for Q9 data.  
(Q1 and Q5 with the telescope at the same orientation also possess
this feature, but the \koic\ difference images, which in the
best of circumstance are low signal-to-noise, were unstable
for these quarters.)
The difference images for
both \koib\ and \koic\ show maxima in the terminal saturation
pixels, rather than in the pixel containing KIC 11295432.
This proves that KIC 11295432 cannot be the source for either
set of transits.
Saturated pixels in these quarters span a rectangle of 
2$\times$4 pixels, or 8\arcsec $\times$ 16\arcsec~ for which centroid
analysis does not rule out background contaminating sources.
Deviations in the difference image along the row direction for 
saturated images likely result from imperfect correction for 
LDE (local detector electronics) undershoot \citep{cald10a},
a signal-dependent offset to negative values along rows traced
to an amplifier in the \ek\ electronics.
The high resolution imaging discussed in
Section 4.2 will provide the primary constraints on potential
background objects, which if these objects were eclipsing binaries could
be the source of false positives.

\begin{figure}
\begin{center}
\includegraphics[height=110mm]{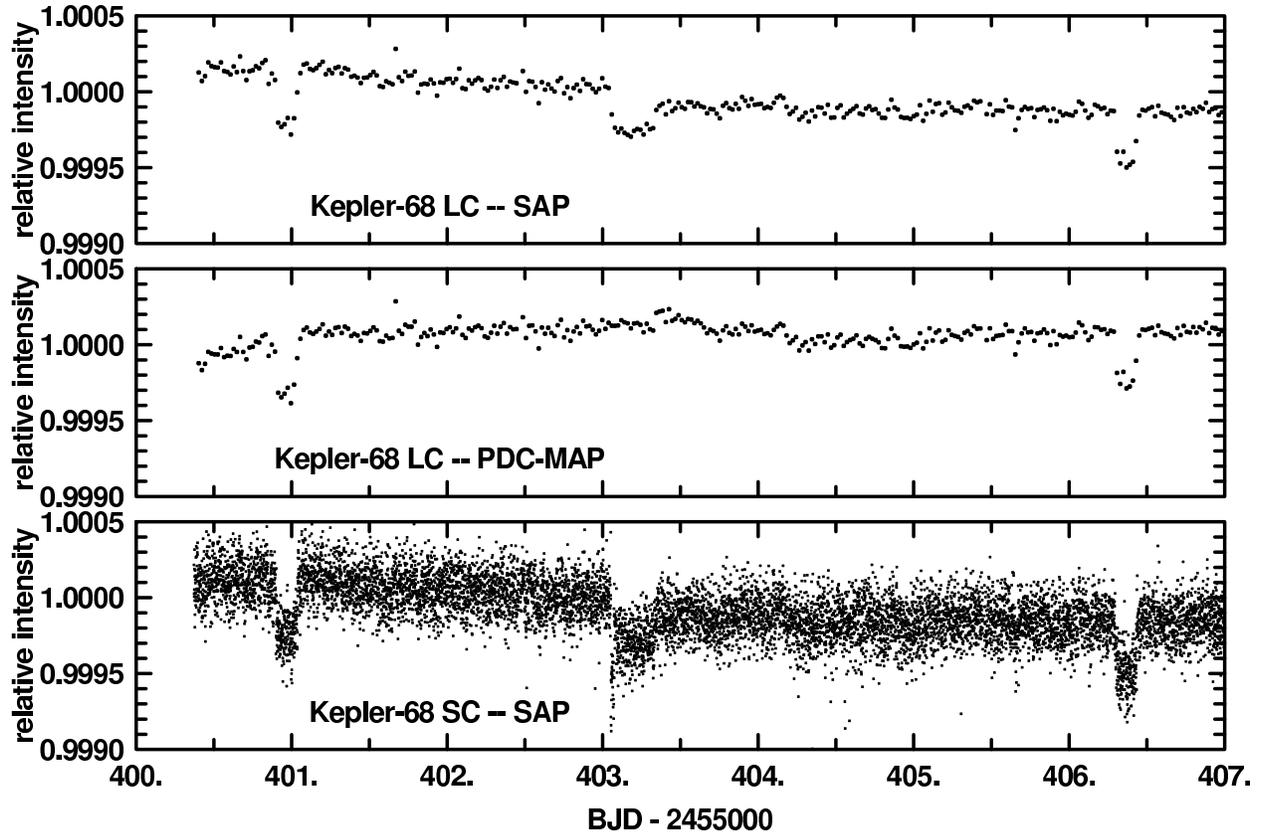}
\end{center}
\caption{The upper panel shows raw \ek\ long cadence data spanning 
six days centered on the ``monotransit" event shown in Fig. 3 of 
\citet{ofir12}.  The middle panel
shows the same data after pipeline processing with PDC-MAP.
The lower panel shows the same time period with short cadence
raw data.}
\label{fig:monoevent}
\end{figure}

In Figure~\ref{fig:monoevent} we present vetting evidence to further
investigate the veracity of the single transit of presumed
long period reported in \citet{ofir12} and shown in their Fig. 3.
The detrending of raw \ek\ data adopted by \citet{ofir12} results 
in a $\sim$8 hour wide intensity drop that is adequately fit with
a standard transit model.  In the raw LC (simple aperture photometry --
no detrending) \ek\ data their event is easily seen at BJD - 2455000 = 403.2.
However, the raw LC data shows a drop of intensity across this 
event of about 0.0001 which may be easily seen by drawing linear
fits to the data before or after the event.  Such behavior is 
commonly seen with sudden pixel sensitivity losses associated
with radiation damage to a single pixel (see, e.g. discussion
from discovery in HST ACS data by \citet{gill07} and \citet{chri12a}
for presence in \ek\ data) -- and ultimate recovery of most, but not all of the 
depressed sensitivity.  Such sensitivity drops can be particularly
difficult to tell from transits in LC data which blur both sudden
(spurious) drops, and short ingress/egress events.  The spurious
signature may be more easily seen in short cadence data as
shown in the bottom panel of Figure~\ref{fig:monoevent} -- here
the event looks more like a sudden sensitivity loss and recovery,
than a transit.  Note that the two occurrences of \koib\ transits
about one day from both ends show symmetric responses about the
transit centers, while the single-event behavior is quite asymmetric
with shape characteristic of a pixel sensitivity loss.
The middle panel shows that the pipeline PDC-MAP treatment 
completely removes any suggestion of the single-event transit
while nicely preserving the real events from \koib.
These considerations do not prove the single-event transit as an instrument
systematic false positive, but this interpretation is
favored by us.  Definitive proof would come if a difference
image analysis across transit could isolate the intensity 
drop to one pixel, while the stellar signal spans several pixels.
We tested for this, but were only able to show that the source
is within the set of saturated pixels for \starname, leaving
open an astrophysical source.

\section{Follow-Up Observations}
\label{sec:fop}

\starname\ was identified as having a candidate planet in late 2009,
prompting follow-up observations to confirm and characterize the planet,
and to secure more precise stellar parameters than are available
in the Kepler Input Catalog.  We were particularly interested in checking
for signs of a nearby eclipsing binary star system that might masquerade
as a planet.  We carried out spectroscopy of the host star \starname \ (Section~\ref{sec:recon})
and high spatial resolution imaging to identify nearby stars in
the photometric aperture (Section 4.2).
Upon passing those gates, we carried out high-resolution,
high signal-to-noise ratio (SNR) echelle spectroscopy with and without an iodine cell to measure atmospheric
stellar parameters, magnetic activity, absorption line shape changes
with time, and to make precise Doppler measurements.  As described below,
these follow-up observations revealed no evidence of a nearby
eclipsing binary for either transiting planet, and they provide
measurements and constraints on the masses of the transiting planets.
In addition, a previously unknown non-transiting planet was revealed
as discussed in Section~\ref{sec:rv}.
Bisector analyses of the high resolution spectroscopy are presented in Section 4.4.
We obtained {\em Spitzer Space Telescope} photometry through transits
of \koib\ and present results in Section 4.5.

\subsection{Reconnaissance Spectroscopy}
\label{sec:recon}

Two reconnaissance spectra were obtained with the Tull Coude Spectrograph
of the McDonald Observatory 2.7m Harlan J. Smith Telescope on the nights
of 25 March 2010 and 28 March 2010, and a third at the TRES Echelle
Spectrograph of the Tillinghast 1.5m telescope on Mt. Hopkins,
also on the night of 25 March 2010.
These high spectral resolution, low signal/noise spectra showed no
convincing evidence for radial velocity variability
at the 0.2\,km\,s$^{-1}$ level, and no hints of any contaminating spectra.
These spectra were cross-correlated against a library of synthetic model
stellar spectra as described by \citet{bata11}
in order to derive basic stellar parameters to compare with the
Kepler Input Catalog (KIC) values.  These spectra were in excellent
agreement, and yielded $\teff = 5750$K, $\logg = 4.0$, and
rotational velocity less than 4\,km\,s$^{-1}$.
The height of the cross-correlation peaks ranged from 0.93 to 0.96,
indicating an excellent match with the library spectra.
This spectroscopy
suggests that \starname\ is a sun-like,
slowly rotating main sequence star, in support of the planetary
interpretation for the transit events. 

\subsection{High Resolution Imaging}

There is always a possibility, especially for bright stars of
high enough proper motion for old plates to reveal the 
background distribution of faint stars in the current epoch.
Inspection of 1953 and 1991 Sky Survey plates shows that 
this does not work for \starname\ -- the proper motion 
is far too small.

\subsubsection{AO Imaging}
\label{sec:ao}

Near-infrared adaptive optics imaging of \starname \ was obtained with the
6.5 m MMT telescope on Mt. Hopkins and the Arizona Infrared imager 
and Echelle Spectrograph (ARIES).  The \starname\ imaging was obtained
on the night of 2010 May 5 (UT) using the f/30 mode with a field
of view of 20\arcsec $\times$ 20\arcsec~ and a resolution of 0{\farcs}02085
per pixel.  The AO system guided on the primary target.  The FWHM
of the $J$ band combined image was 0{\farcs}123, while the $Ks$ band
provided 0{\farcs}112 imaging.  A four-point dither pattern was 
used with a total of 16 images in each band.  Further details of
the MMT-ARIES high resolution imaging for \ek\ follow-up support
may be found in \citet{adam12}.

\begin{figure}
\begin{center}
\includegraphics[height=140mm]{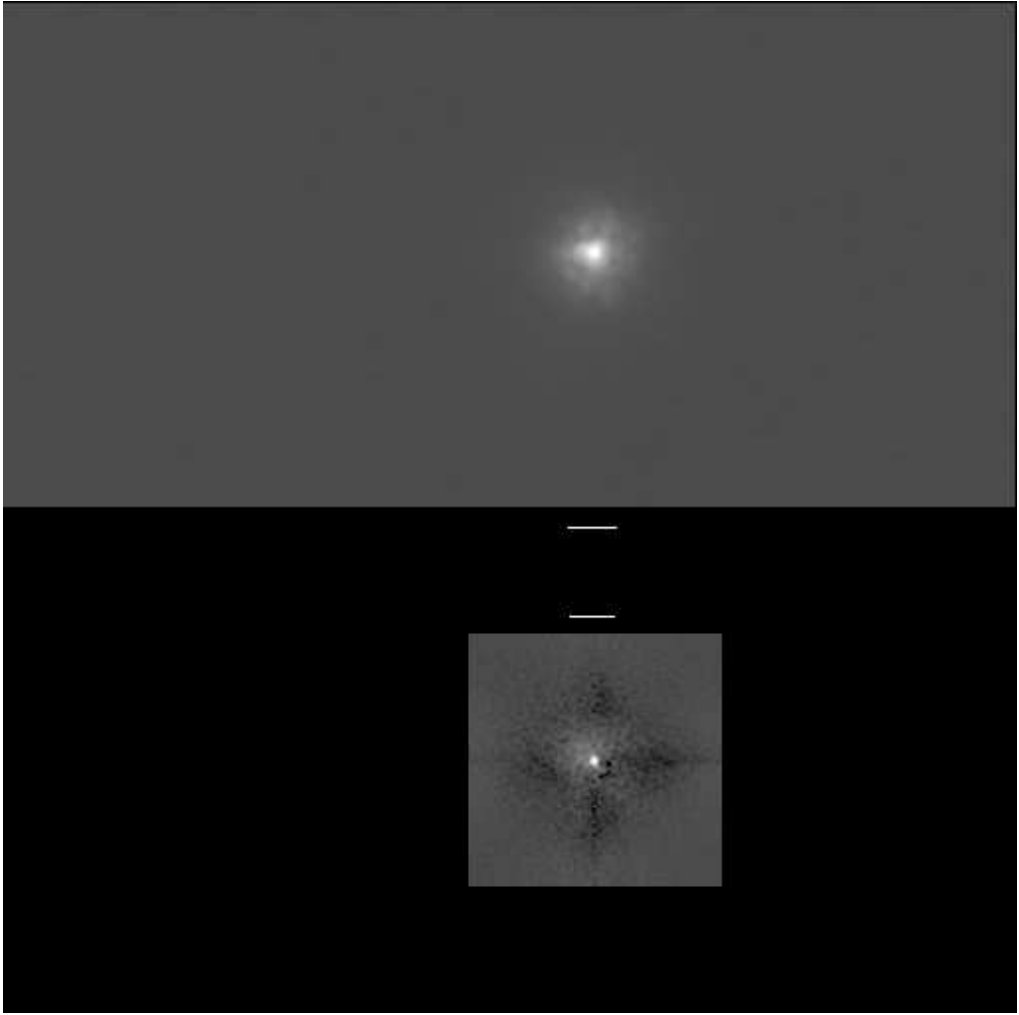}
\end{center}
\caption{The upper panel shows a 10{\farcs}6 by 5{\farcs} region
of the ARIES $Ks$ band AO image.  The lower panel shows the 
full 2{\farcs}76 square R-band Speckle image.  Bars next to 
images illustrate 0{\farcs}5 scale.  Both images have been
normalized to a common central intensity, offset with a 
positive zero point of 1\% of full scale and then displayed
with identical logarithmic stretches.  The speckle image is 
superior for resolution, with the AO being better both in
terms of field of view and limiting depth.}
\label{fig:highres}
\end{figure}

\begin{figure}
\begin{center}
\includegraphics[height=140mm]{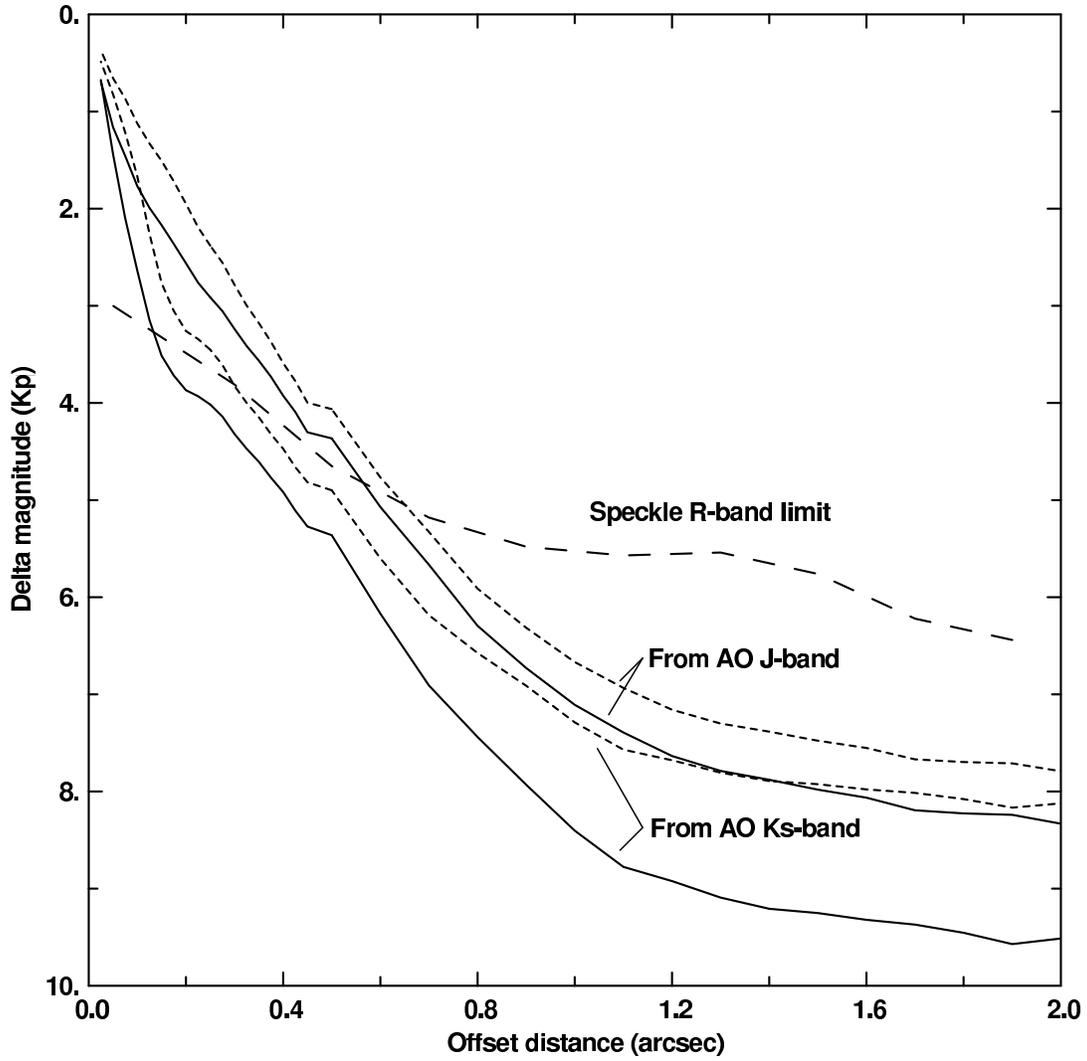}
\end{center}
\caption{Limiting depths provided from the ARIES AO, and 
speckle imaging as a function of offset distance.  
The short dashed lines show the direct $J$ and $Ks$ limits
from ARIES with the associated solid curves being the
estimated $Kp$ limits using the Appendix A transformations
from \citet{howe11}.  The long dashed line shows the 
$R$-band limit from WIYN speckle observations, which has
a central wavelength very similar to the \ek\ bandpass.}
\label{fig:highreslim}
\end{figure}

Figure~\ref{fig:highres} shows a region of the ARIES $Ks$ band
image.  Within the domain plotted no other sources are visible
to the 5-$\sigma$ depth limits as shown in Figure~\ref{fig:highreslim}.
Within the full ARIES field of view one additional source 
at 6.1 and 5.7 magnitudes fainter in $J$ and $Ks$ respectively
was identified.  Using the 2MASS $J$ and $Ks$ magnitudes for \starname\ of
8.974 and 8.587 leads to estimates of 15.07 and 14.29 respectively
for the companion.  Using the transformations in Appendix A of 
\citet{howe11} leads to a $Kp$ = 17.0 $\pm$ 0.4, or a $\delta$-magnitude
of 7.0 $\pm$ 0.4 with respect to \starname\ in the \ek\ bandpass.
This AO identified source also shows up clearly in available 
UKIRT $J$ direct imaging \citep{luca09} and corresponds to the
star KIC 11295432, which does not have a specified $Kp$ in the KIC.
From a combination of \ek\ data centroid consideration (Section 3),
and the ARIES AO, and UKIRT data, 
to roughly the exclusion limits reached in
Figure~\ref{fig:highreslim}, no other sources exist to
relevant larger radii as well.

\subsubsection{Speckle Imaging}
\label{sec:speckle}

Speckle imaging of \starname\ was obtained on the night of 2010 June 20 UT
using the two-color speckle camera at the Wisconsin Indiana Yale NOAO
(WIYN) 3.5-m telescope on Kitt Peak.
The speckle camera simultaneously obtained 1000 40 msec images in two filters:
$V$ (5620/400\AA) and $R$ (6920/400\AA). These data were reduced and
processed to produce a final reconstructed speckle image for each filter.
Figure~\ref{fig:highres} includes this speckle-reconstructed $R$ band image.
The details of the two-color speckle camera are presented in \citet{howe11}
for application in the \ek\ follow-up program.

The speckle data allow detection of a companion star within the approximately
$2.76 \times 2.76$ arcsec box centered on the target.  The speckle imaging can detect,
or rule out, companions between 0.05 arcsec and 1.5 arcsec from \starname.
We found no companion star within the speckle image
to the detection limits shown in Figure~\ref{fig:highreslim}.
When high quality near-IR AO imaging is available, as in this case,
the WIYN speckle imaging is largely redundant providing unique sensitivity
for a small angular separation range due to its superior FWHM
of 0{\farcs}053 -- a factor of two sharper, but generally shallower than the 
MMT AO imaging.

\subsection{Precise Doppler Measurements of \starname}
\label{sec:rv}

We obtained 52 high resolution spectra of \starname\ between 2010 July 19
and 2012 Aug 12 using the HIRES spectrometer on the Keck~I 10-m
telescope \citep{vogt94}.  We configured the spectrometer,
took observations, and reduced the spectra with the same method
used on thousands of nearby FGKM stars \citep{marc08}.
This technique yields
a Doppler precision of 1.0-1.5 \ms\ for stars as faint as 14th magnitude (V band),
depending on spectral type and rotational \vsini.
The HIRES fiber-feed was not used for these observations.
An iodine cell was used to superimpose iodine lines
on the stellar spectrum, providing empirical information for
each exposure and each wavelength about the
instantaneous wavelength scale and instrumental profile of the spectrometer.  

The observations were made with the ``C2 decker'' entrance aperture
which projects to $0\farcs87 \times 14\farcs0$, giving a
resolving power of about 60,000 at 5500 {\AA} and enabling sky
subtraction (typical seeing is $0\farcs6 - 1\farcs2$).    The
average exposure was 11 minutes, giving a signal-to-noise ratio per pixel of 200.  

\begin{figure}
\begin{center}
\includegraphics[width=150mm]{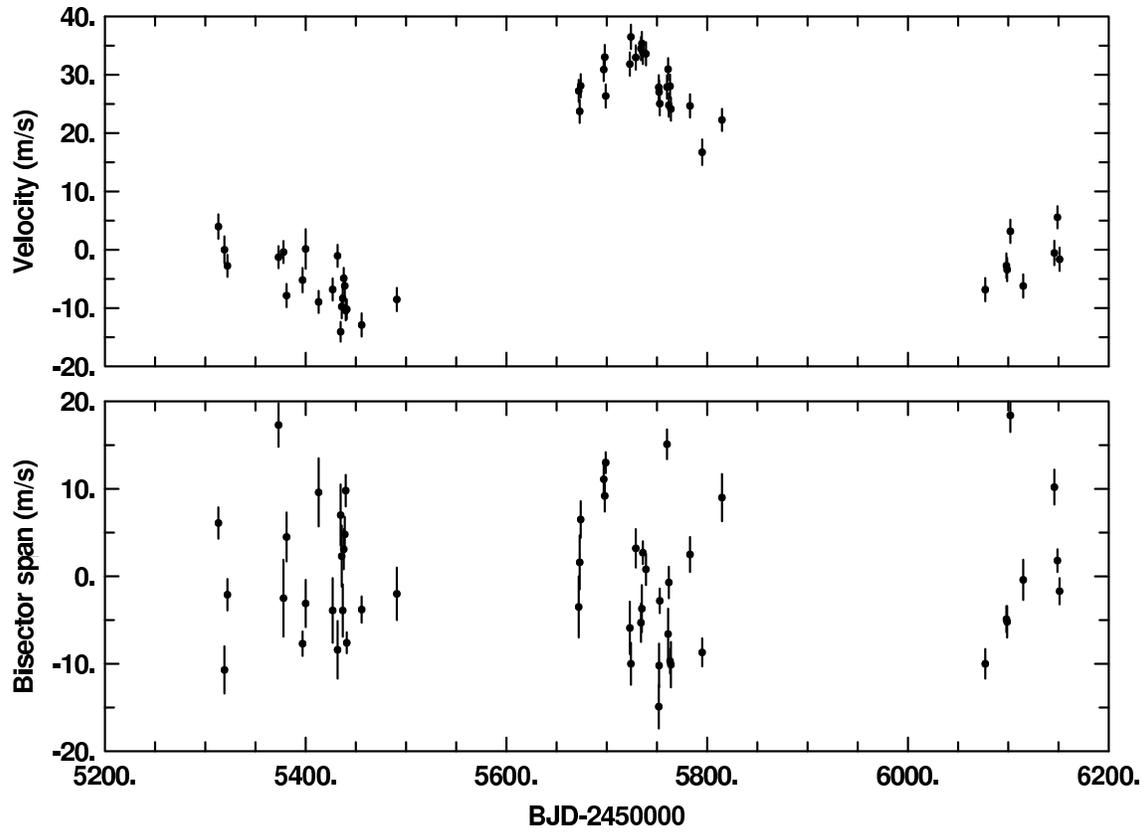}
\end{center}
\caption{Radial velocities versus time from the Keck-HIRES
spectra over 2010-2012 are shown in the upper panel.
Error bars include the internal
uncertainties and the expected combined astrophysical and
instrumental jitter of 1.5 \mse, added in quadrature.
The lower panel shows line bisectors derived from the 
same spectra.}
\label{fig:rv_time}
\end{figure}

The raw CCD images were reduced in the standard way, including
the subtraction of background sky counts (mostly from moonlight)
at each wavelength just above and below the
stellar spectrum.  We used Doppler analysis with the algorithm of \citet{john09}.  The
internal Doppler errors (the weighted uncertainty in the mean of 400
spectral segments) are typically 1.0-1.5 \ms.   Our experience
with hundreds of G-type main sequence stars shows that the actual
errors are larger than the internal errors by $\sim$1.5 \ms for
such stars.  Thus, we added 1.5 \ms in quadrature to the internal
uncertainties to yield our estimated uncertainty in the RV.   The resulting
velocities and uncertainties are given in Table~\ref{tab:radtab}
and shown in Figure~\ref{fig:rv_time} as a function of time.
The error bars include the internal Doppler errors and an assumed
jitter of 1.5 \ms\ . 

The center of mass velocity relative to the solar system barycenter
(Gamma Velocity) for \starname\ is
$\gammaVelb$ km s$^{-1}$ (Table~\ref{tab:SystemParams}).
This is a typical radial velocity for a star in the Galactic disk,
indicative of a middle-aged disk star.  The near solar
metallicity (Section~\ref{sec:spectParam}), $\feh=\fehsys$,
magnetic activity (Section~\ref{sec:spectParam}), and
asteroseismic age, $\ageKASC$ Gyr (Section~\ref{sec:astero}),
also suggest Galactic disk membership. 

\begin{figure}
\begin{center}
\includegraphics[width=85mm]{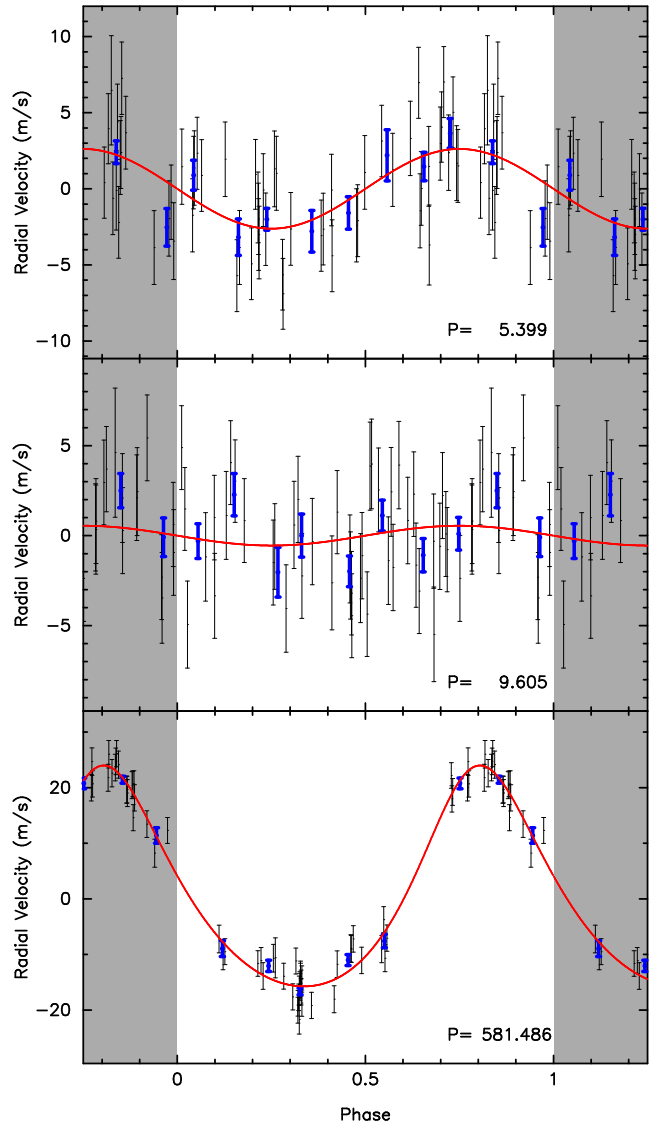}
\end{center}
\caption{This figure shows the radial velocity measurements and model fits. 
The top panel shows the RV measurements and model phase folded to the 
orbital period of \koib.  The black lines show the radial velocity
measurements and 1-$\sigma$ uncertainties after the removal of the best 
fit model for \koic\ and \koid.  The thick blue points show the same
RV data but averaged in 0.1 phase bins.  The red line is the best fit 
Keplerian orbital model.  The orbital period is indicated in the lower 
right portion of the panel.  The middle and bottom panels show the RV 
measurements in similar fashion for \koic\ and \koid\ respectively.
The fits for \koib\ and \koic\ have eccentricity forced to zero,
while the \koid\ solution allowed this as a parameter.}
\label{fig:rv3pan}
\end{figure}

The primary signal from the RVs is the $K$ = 19.9 $\pm$ 0.75 m s$^{-1}$
variation with a period of 580 $\pm$ 15 days as shown first in 
Figure~\ref{fig:rv_time}, and phased in Figure~\ref{fig:rv3pan}.
This clear RV signature, coupled with the lack of bisector
variations discussed in Section 4.4 provides discovery and
confirmation of the Jovian-scale outer planet -- \koid.

The velocities phased to the photometric period of \koib\
in Figure~\ref{fig:rv3pan} show a
clear, continuous, and nearly sinusoidal variation consistent
with a nearly circular orbit of
a planetary companion.
Note that the 7 magnitude (see Sections 3 and 4.2.1) fainter
companion KIC 11295432 blended into the \ek\ photometry does
not fall within the HIRES slit and cannot be the source 
of \koib\ radial velocity variations.
The lack of any discontinuities in the phased
velocity plot argues against
a background eclipsing binary star as the explanation.  Such a binary
with a period of 5.4 d would have orbital semi-amplitudes of tens of
kilometers per second, so large that the spectral lines would
completely separate from each other, and separate from the lines of the
main star.   Such breaks in the spectral-line blends would cause
discontinuities in the velocity variation, which are not seen here.
Thus, the chance that the 5.4 d periodicity exhibited independently
in both the photometry and velocities might be caused by an eclipsing
binary seems small.

Precision Doppler measurements are used to constrain the mass
of \koib\ as discussed in Section~\ref{sec:planet}.
The absence of a Doppler signal for \koic\ is used to compute an
upper limit to the mass under the planet interpretation.

\subsection{Bisector Analysis}
\label{sec:bisector}

From the Keck spectra, we computed a mean line profile and the
corresponding mean line bisector. Time-varying line asymmetries are
tracked by measuring the bisector spans -- the velocity difference between
the top and bottom of the mean line bisector -- for each spectrum \citep{torr05}.
When radial velocity variations are the result of a blended
spectrum between a star and an eclipsing binary, we expect the bisectors
to reveal a phase-modulated line asymmetry \citep{quel01, mand05}.

The bisector spans in the lower panel of Figure~\ref{fig:rv_time} show no correlated
variation with the radial velocities and have a scatter of 7.8 \mse,
which is significantly less than the semi-amplitude of \koid.
However, the uncertainties in the bisector measurements are larger than the
semi-amplitude of the two smaller planets in the system and a bisector
analysis is thus inconclusive with respect to \koib\ and \koic.

\subsection{Photometry with the Spitzer Space Telescope}

\koib\ was observed during two transits with \wspitzer/IRAC 
\citep{werner04,fazio04} at 4.5~\micron\ (program ID 60028). 
The observations occurred on UT 2010 December 27
and on UT 2011 January 7. The visits lasted 7.6~hours. 
The data were gathered in subarray mode ($32\times32$ pixels)
with an exposure time of 2~s per image which yielded 15680 images per visit. 

\begin{figure}
\begin{center}
\includegraphics[height=140mm]{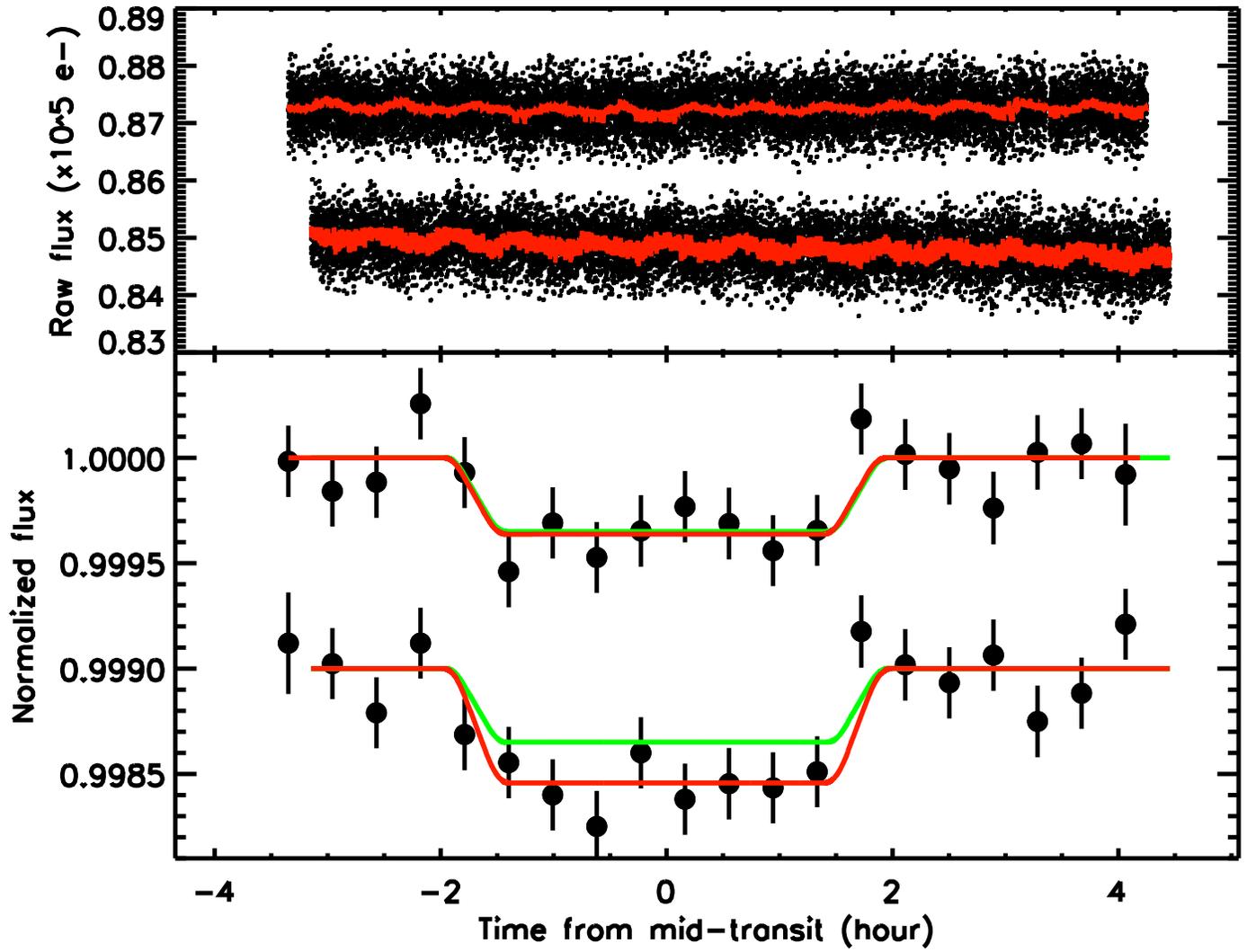}
\end{center}
\caption{\spitzer\ transit light-curves of \koib\ observed in
the IRAC band-pass at 4.5~\micron. Top panel~: raw (unbinned)
transit light-curves for the two visits of  \koib.
The second visit (at the bottom) is shifted vertically from the
first visit (at the top) for display purpose. The red solid
lines correspond to the best fit model which include the time and
position instrumental decorrelations as well as the model
for the planetary transit (see text). Bottom panel~: corrected,
normalized, and binned by 23 minutes transit light-curves
with the transit best-fit plotted in red and the transit shape
expected from the \ek\ observations overplotted as a green line.
The second visit has again been shifted down for display.}
\label{fig:spitzerlightcurves}
\end{figure}

The images used are the Basic Calibrated Data (BCD)
delivered by the \emph{Spitzer} archive.
These files are corrected for dark current, flat-fielding,
detector non-linearity and converted into flux units.  
The method we used to produce photometric time series from
the images is described in \cite{desert09}.
We first discard the first half-hour of observations,
which are affected by a significant telescope jitter before stabilization. 
To facilitate the evaluation of the photometric errors,
we then convert the pixel intensities to electrons using the
information given in the detector gain and exposure time provided in the FITS headers.
We convert to UTC-based BJD following the procedure developed by \citet{eastman10}.
We correct for transient pixels in each individual image
using a 20-point sliding median filter of the pixel intensity versus time.
We find the centroid position of the stellar point spread function
(PSF) and perform aperture photometry using a circular aperture
with a radius of $3.5$~pixels on individual BCD images;
we adopt the radius which provides the smallest errors.
The final number of photometric measurements used is $13146$
data points for the first visit and $13262$ for the second one.
The raw time series are presented in the top panels of Figure~\ref{fig:spitzerlightcurves}.
We find a typical signal-to-noise ratio (SNR)
$260$ per image which corresponds to about 90\% of the theoretical signal-to-noise.

We used a transit light curve model multiplied by instrumental
decorrelation functions to measure the transit parameters and their
uncertainties from the \spitzer\ data as described in \cite{desert11a}. 
We compute the transit light curves with the IDL transit routine
\texttt{OCCULTSMALL} from \cite{man02}.
In the present case, this function depends on one
parameter: the planet-to-star radius ratio $R_p / R_\star$.
The other transit parameters are fixed to the value derived from the \ek\ lightcurves.
The limb-darkening coefficients are set to zero consistent with
expected small values in the IR and insufficient photometric 
precision in the \spitzer\ light curves to matter.

The \spitzer/IRAC photometry is known to be systematically affected by the
so-called \textit{pixel-phase effect} (see e.g., \citealt{charbonneau05,knutson08}).
We decorrelated our signal in each channel using a linear
function of time for the baseline (two parameters) and a
quadratic function of the PSF position (four parameters) to correct the data for each channel.
We performed an MCMC analysis with six chains of length $10^5$ each
providing median depth values and errors.
We allow asymmetric error bars spanning $34\%$ of the points
above and below the median of the distributions to derive the $1~\sigma$
uncertainties for each parameter as described in \citet{desert11b}.

We measured the transit depth for \koib\ at 4.5~\micron\ of
350 $\pm$ 70 ppm for the first
and 560 $\pm$ 70 ppm
for the second visit.  The weighted mean of these two values
provides a transit depth of 455 $\pm$ 50 ppm.
The value for the first visit is in excellent agreement with the \ek\ depth
of 346 ppm suggesting that the radius-ratio of the candidate \koib\
to its host star is a wavelength independent function,
in agreement with a dark planetary object.
However, visit two is 3 $\sigma$ from the \ek\ value.
Two possibilities for this behavior, other than the
small statistical probability of two such discrepant values
legitimately following from the same underlying distribution,
are: (1) a physically different depth in the two epochs for the
transit, and (2) an inconsistency in one of the \spitzer\ epochs
that we have failed to uncover.  To investigate the first possibility
we identified the position of the two \spitzer\ epochs relative to
\ek\ coverage.  Alas, despite the excellent overall duty cycle with
\ek\ in excess of 90\%, both epochs fell into the longest downtime
experienced to date with \ek\ -- due to a safing event before Q8.
We have no data with which to challenge the hypothesis that the
two epochs really do have different depths, as could happen if 
there is an as yet unclaimed transiting planet that overlapped
with the deeper \spitzer\ transit of \koib.  We have not 
uncovered direct evidence for an inconsistency in the data 
quality or analyses sufficient to explain the discrepancy in 
depths of the two \spitzer\ epochs.  However, we do note that
the scatter evident in Figure~\ref{fig:spitzerlightcurves} for 
the anomalously deep transit is much larger than we normally 
encounter.

Given the perfect agreement of one \spitzer\ visit in reproducing
the transit depth seen in the optical, the radial velocity 
confirmation documented in the previous section, and a strong
\blender\ validation to be presented in Section 5 the case
for the planet interpretation of \koib\ remains strong.

\section{\blender \ analysis of the \ek\ light curve}
\label{sec:blender}

In this section we examine the possibility that the transit signals
seen in the \ek\ photometry of \starname\ are the result of
contamination of the light of the target by an eclipsing object along
the same line of sight. We begin with the more difficult signal
\koic, which would correspond to an Earth-size planet.

\subsection{Validation of \koic}

In the absence of a dynamical confirmation of the planetary nature of
the \koic\ signal from either a Doppler detection or a Transit Timing Variation (TTV)
signature, we proceed here with a probabilistic ``validation''. In
essence, we seek to demonstrate that the signal is much more likely to
be caused by a bona-fide transiting planet than by a false positive
(or ``blend'').  For this we applied the \blender\ procedure, which
has been described previously \citep{torr04, Fressin:11,
Fressin:12a} and used to validate a number of other \ek\ planets
\citep[e.g.,][]{Ballard:11, Lissauer:12, Borucki:12, Howell:12,
Gautier:12}. We refer the reader to these works for full details of
the method.

Briefly, we performed a systematic exploration of the different types
of false positives that can mimic the signal, by generating large
numbers of synthetic blend light curves over a wide range of
parameters and comparing each of them with the \ek\ photometry in
a $\chi^2$ sense. The photometry we used for the validation is the
de-trended short-cadence time series (one month from Q2, plus
the seven quarters of Q5--Q11 -- 886,638
measurements), which provides stronger constraints on the shape of the
transit than the long cadence data, as shown later.  We rejected
blends that result in light curves inconsistent with the \ek\
observations.  We then estimated the frequency of the remaining blends
by taking into account all available observational constraints from
the follow-up observations mentioned above. Finally we compared this
frequency with the expected frequency of true planets (planet
``prior'') to derive the odds ratio.

The types of false positives we considered include eclipsing systems
falling within the \ek\ aperture that are either in the background
or foreground, or that are physically associated with the target in a
hierarchical triple configuration. We allowed the object producing the
eclipses to be either a star or a planet. To compute the blend
frequencies we used informed estimates of the number density of stars
in the background from the Besancon Galactic structure model of
\cite{Robin:03}, rates of occurrence of eclipsing binaries in the
\ek\ field from the work of \cite{Slawson:11}, and frequencies of
planets involved in blends based on the catalog of planet candidates
(KOIs) of \cite{bata12}, which was constructed using observations
from Q1 to Q6. The same catalog was also used to estimate the planet
prior. This list of KOIs is bound to contain some fraction of false
positives \citep[see, e.g.,][]{Morton:11, Morton:12} and it is also
likely incomplete mainly due to the difficulty in detecting shallow
signals especially with long periods. Consequently we have applied
corrections for these effects following a Monte Carlo procedure
described by \cite{Fressin:12b}, both in computing the blend
frequencies and also for the planet prior.

The observational constraints used to further reduce the number of
blends included the following: (a) the color of the star as reported in
the KIC \citep{Brown:11}, which allows us to rule out any simulated
blends resulting in a combined color that is significantly redder or
bluer than the target; (b) limits from the centroid motion analysis on
the angular separation of companions that could produce the signal
(Section\ 3); (c) brightness and angular separation limits from
high-resolution adaptive optics and speckle imaging (Section\ 4.2); (d)
limits on the brightness of unresolved companions from high-resolution
spectroscopy (Section\ 4.1).
For eclipsing systems physically associated with the
target we considered also dynamical stability constraints in
hierarchical triple configurations \citep{Holman:99}.

The \blender\ simulations for \koic\ rule out all false positive
scenarios involving eclipsing binaries physically associated with the
target, as the predicted light curves invariably have the wrong shape
to match a planetary transit.  For the scenarios involving eclipsing
binaries that are in the background or foreground we find a blend
frequency of $2.8 \times 10^{-6}$, and for those in which larger
planets transit background stars we estimate a much smaller frequency
of $7.0 \times 10^{-8}$. Hierarchical triples (a larger planet
transiting a stellar companion to the target) contribute a frequency
of $6.7 \times 10^{-7}$.  The total blend frequency is thus $3.5
\times 10^{-6}$.  An illustration of the constraints on false
positives resulting from \blender\ as well as those from the follow-up
observations is given in Figure~\ref{fig:blender246.02}.

An estimate of the planet prior may be obtained by dividing the number
of known planets of similar size as \koic\ from \cite{bata12}
by the total number of main-sequence \ek\ targets observed during
Q1--Q6, which is 138,253. We find 71 KOIs that are in the same
(3$\sigma$) radius range as the putative planet in \koic, of
which we expect 9.4 to be false positives, following the procedures of
\cite{Fressin:12b}. We also expect such shallow ($\sim$50 ppm) transit
signals to be detectable around only about 9.7\% of all \ek\
targets, which we use to correct for incompleteness. We compute the
planet prior as $(71 - 9.4)/0.097/138,\!253 = 4.6 \times 10^{-3}$.
This \emph{a priori} planet frequency is $4.6 \times 10^{-3}/3.5
\times 10^{-6} \approx 1300$ times larger than the estimated blend
frequency, from which we conclude that \koic\ is validated as a
true planet with a very high degree of confidence.

In the above calculations we have not explicitly taken into
consideration the period of the signal, which may be an important
factor for small candidates such as \koic\ because such signals
are relatively rare in the KOI list of \cite{bata12}.  This may in
principle influence both the planet prior and the blend frequencies we
have just described, given that the latter also draw on the KOI list
to estimate the rate of occurrence of larger planets involved in
blends. Therefore, instead of allowing eclipsing binaries and
transiting planets with any orbital period for the blend frequency
calculations, we repeated the analysis with the more realistic
approach of accepting only blends with periods near the measured
period of \koic\ (within a factor of two). We did the same when
computing the planet prior, for consistency. The total blend frequency
is reduced in this way by about a factor of five, but the new planet
prior is only 1.5 times smaller, resulting in a larger net odds ratio
of $\sim$4300 that provides for an even stronger validation than
before.

The above calculations neglect the fact that KOI-246.02 was found around 
a target that has a statistically validated transiting planet, KOI-246.01 = 
Kepler-68b.  The planet priors used were averaged between  
single and multi-planet systems.
Actual planet priors are about 30\% smaller for single planets and more
than an order of magnitude larger for multiple planets \citep{Lissauer:12, Liss13}.
The presence of a known planet also
increases the prior for physically-associated blends, but by a  smaller 
factor. The numbers quoted above for the likelihood of false positives are four 
times higher for background blends than for physically-associated 
blends, so when multiplicity of the system is accounted for, the odds 
ratios against blends quoted above are increased by roughly an order of 
magnitude.

As an interesting test of the value of short cadence versus long
cadence data for false positive discrimination, we repeated the
calculations with the long-cadence time series (Q1--Q2, Q4--Q11,
34,556 measurements) including the period cut described above. We find
an odds ratio of approximately 1500, about three times lower than when
using short cadence, but still large enough to comfortably validate
the signal. Therefore, at least for \koic, short cadence provides
a clear advantage in ruling out blends. This is likely due to the
better definition of the transit shape at ingress/egress, which is
often where the main differences are between model light curves for
blends and the model for a true planet. In this case we find that the
main improvement (decrease) in the blend frequencies when going from
long to short cadence is in the scenarios involving hierarchical
triples, with background eclipsing binary frequencies changing the
least.

While in principle the above calculations provide a clear statistical
validation of the \koic\ signal as a true planet, independently of
the detection of the reflex motion of the star (radial velocities), we
note that some of the blend scenarios involving background eclipsing
binaries yield a fit to the \ek\ photometry that is significantly
\emph{better} than that of a planet fit. This is a situation we have
not encountered in previous validations of \ek\ candidates. The
light curves of these false positive scenarios all feature a very
shallow ($\sim$10 ppm) secondary eclipse that happens to match a
similar dip present in the phase-folded photometry for \koic\ at
phase 0.5.
This shallow dip at phase 0.5 has a formal significance of 3 -- 4 $\sigma$,
and is the primary source for these 
favorable false positive fits.

  The reality of the signal at phase 0.5 for \koic\ is in doubt;
were this not so claiming validation in this case would be impossible.
Figure~\ref{fig:sececlipse} contrasts phase folded short cadence data
for \koic\ at both the transit, and offset by half phases.
In assessing the reality of this putative secondary eclipse feature
we have examined the data in several ways.  When averaged over 
widths comparable to the transit the data show a few
other excursions (both positive and negative) to deviations
as large as the phase 0.5 feature; thus the feature is obviously not
highly significant.  However, other aspects of data investigation do
not support a straightforward dismissal of evidence consistent with
the weak, $\sim$10 ppm feature.  Examination of medians over the 
phase bins shown in Figure~\ref{fig:sececlipse} showed the same offsets,
thus the deviation does not result simply from a small subset of the data.
Likewise, dividing the data into first and second halves before doing
independent phase binnings shows evidence for a $\sim$10 ppm depression
at phase 0.5 in both cases.  In these two halves there are again other
instances of deviations of comparable width and depth; however, across
the two halves the only cases lining up are those at phase 0.5.
While nothing here is convincing in terms of regarding the phase 0.5
offset as real, it is equally the case that  
the reality of the offset cannot be excluded.
We therefore took the conservative approach
of allowing \blender\ to be influenced by the apparent negative offset
at phase 0.5, which still provides the favorable odds ratio required
for formal validation of \koic\ as a planet.

Further complicating this interpretation these data show a small
degree of correlated noise, even after folding over 64 orbits of \koic.
After binning to 0.001 phase bins the first autocorrelation value 
is 0.25, falling to zero within about 0.005 in phase.  We formed 
simulated time series consisting only of Gaussian noise, stellar 
oscillations, and granulation \citep{gill11b} appropriate to \starname\
and found that after folding this had a similar (although smaller)
degree of autocorrelation suggesting that inherent stellar variations
may suffice to explain the modest correlated noise.

\subsection{Validation of \koib}

The robust detection of a Doppler signal at the period of \koib\
provides strong support for the planetary nature of that signal. There
is, however, a small chance that a background eclipsing binary could
mimic this spectroscopic signature of a planet (as well as the
photometric signal), although this possibility was convincingly 
argued against in Section 4.3.  The precision of our bisector span measurements
in Section\ 4.4 is not quite high enough for a definite conclusion
regarding this possibility, so we proceed here with a validation
analysis along the lines of what was done for \koic.

The much deeper transit ($\sim$350 ppm) and higher signal-to-noise
ratio of \koib\ result in significantly stronger constraints on
the shape of the signal, and consequently in a much reduced frequency
of blends that give acceptable fits to the \ek\ light curve.
Indeed, background eclipsing binaries are completely ruled out by
\blender, as no such scenarios yield light curves with a transit shape
that matches the observations sufficiently well.  And while some
hierarchical triple scenarios are allowed by \blender, the companion
stars would all be bright enough that they would have been seen in our
high-resolution spectra. The only remaining blend scenarios that are
viable are those involving larger planets transiting a background or
foreground star.  We estimate the frequency of such blends to be $1.4
\times 10^{-8}$.

To estimate the a priori likelihood of a true planet we use the fact
that there are 96 candidates in the list of \cite{bata12} with a
similar planetary radius as that implied by \koib\ (within
3$\sigma$), of which an estimated 6 should be false positives. Signals
of this kind are expected to be detectable in 58.7\% of all \ek\
targets. The planet prior is then $(96 - 6)/0.587/138,\!253 = 1.1
\times 10^{-3}$. With this we obtain an odds ratio of nearly 79,000 in
favor of the planet interpretation, i.e., a very clear validation of
\koib.  The above results used the long-cadence time series for
simplicity, and included the period cut described above. Use of short
cadence would likely result in an even higher odds ratio.
For completeness we note that \citet{Morton:11} reported 
a false positive probability of 0.01 for \koib, whereas 
our confidence level is orders of magnitude better.
Their result was based on a less sophisticated analysis,
much less \ek\ data, and didn't make use of any of the 
follow-up observations that we have utilized.

\section{Stellar Characteristics}
\label{sec:star}

\subsection{Spectroscopic Parameters}
\label{sec:spectParam}

We did a spectral synthesis analysis using SME \citep{vale96,
vale05} of one of our high resolution template spectra from Keck-HIRES of
\starname\ to derive an effective temperature, $\teff  = \teffSMEOrig$
K, surface gravity, $\logg = \loggSMEOrig$ (cgs), metallicity, $\feh = \fehSMEOrig$, 
and $\vsini = \vsiniSMEOrig$ \kms.

The above effective temperature was used to constrain the
fundamental stellar parameters derived via asteroseismic analysis
(see Section~\ref{sec:astero}).  The asteroseismology analysis
gave log $g$ = $\loggKASC$ which is 0.1 dex higher than the SME value.
The asteroseismology value is likely superior because of the
high sensitivity of the acoustic periods to stellar radius.
Still, the asteroseismology result depended on adopting the
value of $\teff$ from SME.  We recomputed the SME analysis by
freezing (adopting) the seismology value for $\logg$.
See \citet{torr12} for a recent discussion of such iteration
in the analogous context of high SNR transit light-curve
analysis providing the $\logg$ ``truth".
This iteration yielded values of $\teff = \teffSME$ K, $\feh = \fehSME$,
and rotational $\vsini = \vsiniSME \kms$.   The revised effective
temperature was then put back into the asteroseismology calculation
to further constrain the stellar radius and gravity.   This iterative
process converged quickly, as the \logg\ from seismology yielded an
SME value for \teff \ that was only slightly different from the
original unconstrained determination, and the
asteroseismic log $g$ using the iterated $\teff$ remained at 4.281.
Likely systematic errors on \teff \ and \feh \ of 59 K and 0.062 dex
have been derived by \citet{torr12} in comparing results across
multiple spectroscopic packages for a large number of stars.
Adding these additional errors in quadrature with the errors
quoted above results in more resonable total errors of 74 K and 0.074 dex.

We measured the Ca II H\&K emission \citep{isaa10}, yielding
a Mt. Wilson S value, $S$=0.139 and $\log R'_{\rm HK}$ = -5.15.
These values suggest lower activity for \starname\ in comparison to
mean solar reported as
$S$ = 0.178, $\log R'_{\rm HK}$ = -4.90 by \citet{lock07} and
$S$ = 0.171, $\log R'_{\rm HK}$ = -4.96 by \citet{hall09}.
Our own measure of solar activity using Ganymede as a proxy
yielded $S$ = 0.164 and $\log R'_{\rm HK}$ = -4.97 \citep{isaa10}
for 2009 August 31 when the Sun was still at a low state of its cycle.
Thus \starname\ is a magnetically inactive star, consistent with
its low rotational rate, \vsini = 0.5 \kms.   \starname\
appears to be a middle-aged (age 2-10 Gyr) slowly rotating
inactive star, on the main sequence.  This is consistent with
the age derived from the asteroseismology analysis (Section~\ref{sec:astero}).

The activity indices, $S$ and $\log R'_{\rm HK}$, have modestly lower than solar values,
while Figure~\ref{fig:q4phot} suggests significantly lower than typical levels
of photometric variability compared to the Sun.
\citet{hall09} summarize 14 years of contemporaneous photometric
and spectroscopic (for activity) measurements of 28 solar analog
stars with precisions in the photometry capable of detecting 
changes at roughly half the level seen for the Sun.
Although a strong correlation between photometric variability
exists with activity, they find that stars with near-solar
activity indices show a range of half (not well bounded) to
twice solar in photometric variability.  \starname\ seems 
consistent with this lower range of photometric change at a 
given activity level, although not knowing what phase of a
\starname\ activity cycle was sampled limits fidelity.
Multi-year \ek\ data will eventually enable robust activity -- 
photometric variability understanding.

\subsection{Asteroseismology and the Fundamental Stellar Properties}
\label{sec:astero}

%\note [Mass, radius, luminosity, surface gravity, age]

The utility of asteroseismology for exoplanet interpretations fundamentally
rests on recognizing that both asteroseismic and transit light curve 
modeling (at high signal to noise levels and with known or
assumed eccentricity) provide constraints on the
same stellar parameter -- \rhostar.  Thus high precision knowledge
of \rhostar\ which commonly results when asteroseismology is feasible
provides for a natural means of tightening the exoplanet transit light
curve solution by adopting this as a prior -- see \citet{gill11a} and 
\citet{nutz11}.  \starname\ at $Kp$ = 10.00, near solar temperature, and a
KIC radius of 1.06 $R_{\odot}$ was recognized early as an excellent
candidate for asteroseismology with a prediction of 99\% chance of 
success with only two months of short cadence data \citep{chap11}.
\starname\ was observed at Short Cadence (SC) \citep{gill10}
for one month in Q2 as a KASC survey target,
then added to the science team SC targets from Q5 onwards to support
asteroseismology and fine analyses of the short transit ingress and egress.

The power of asteroseismology in setting estimates of the stellar 
radius, which determines the exoplanet radius via ratio with the 
transit depth depending on the square of this ratio, can be seen by
recounting knowledge of $R_*$ for \starname\ at the time of initial analyses.
Transit light curve solutions for \koib, with stellar radius as a free
parameter returned values of 1.63 $R_{\odot}$, compared to a KIC radius of 1.06 $R_{\odot}$
(with spectroscopic solutions favoring something like the KIC value).
These two stellar radius values would lead to differences of a factor
of 3.6 in the planetary density emphasizing the need for
better knowledge of the stellar radius.

Figure \ref{fig:power} shows the \starname\ power spectral density with
input of 10 months of SC data, the near-evenly spaced peaks characteristic
of solar-like oscillations are obvious.  Even when first inspecting
power spectra from single months of SC data the spacing between modes
was trivial to estimate at about 100 $\mu$Hz compared to a solar value
of 135 $\mu$Hz.  Since the mean stellar density is known to scale
as the square of frequency spacing \citep{ulri86}, this allowed an early constraint
that \starname\ was at 0.55 of the solar mean density, and assuming a solar
mass (given the solar \teff) 
provided an estimate of 1.22 $R_{\odot}$ at the back-of-the-envelope
level of analysis.

The high SNR of the power spectrum shown in Figure~\ref{fig:power}
made peak-bagging (derivation of frequencies for individual modes)
straightforward, and eight team members fit the modes using 
Maximum Likelihood Estimation (MLE), and MCMC approaches 
(e.g., see \citet{flet09}; \citet{hand11}; \citet{appo12}).
The adopted frequencies listed in Table~\ref{tab:freqs} came from
the single set that most consistently was at the median over all
eight.

The fitting technique has been reported in various versions 
for the analysis of {\em HST} observations of HD 17156 \citep{gill11a},
and \ek\ observations of HAT-P-7 \citep{jcd10} and Kepler-10 \citep{bata11}.
The underlying stellar evolution modelling is provided using the 
ASTEC code \citep{jcd08a} with eigenfrequency analyses coming
from ADIPLS \citep{jcd08b}.

In the present case the stellar parameters grid 
includes a few values of the mixing-length parameter
$\alpha_{\rm ML}$
%(for relatively low-mass stars) or the overshoot parameter 
%$\alpha_{\rm ov}$  (for somewhat higher-mass stars)
in addition to the mass $M$ and the initial composition,
characterized by the abundances $X$ and $Z$ by mass of hydrogen and
heavy elements.
Thus the evolution sequences are characterized by a set of
parameters $\{\CP_k\} = \{M, Z, X, \alpha_{\rm ML}\}$. %, \alpha_{\rm ov}\}$.
Details of the \starname\ grid are presented below.

The fit of a given model to the data is defined in terms of 
\begin{equation}
\chi^2_\nu = {1 \over N-1} \sum_{i=1}^N 
\left( {\nu^{\rm (obs)}_i - \nu^{\rm (mod)}_i \over \sigma_i } \right)^2 \; ,
\label{eq:chisqnu}
\end{equation}
where $\nu^{\rm (obs)}_i$ and $\nu^{\rm (mod)}_i$ are the observed
and model frequencies and $\sigma_i$ is the error in the observed
frequencies.
It is assumed that the degree and order of the observed frequencies have
already been determined.
In addition, an augmented fit

\begin{equation}
\chi^2 = \chi^2_\nu + 
\left( {T_{\rm eff}^{\rm (obs)}- T_{\rm eff}^{\rm (mod)} 
\over \sigma(T_{\rm eff}) } \right)^2  \; ,
\label{eq:chisq}
\end{equation}
including the observed and model effective temperature $T_{\rm eff}$ is formed.

For each evolution sequence frequencies are calculated for
selected models along the sequence (typically every fifth),
but such that frequencies are also available at all models in the sequence
in the vicinity of the model $\CM_{\rm min}^\prime$ 
with the smallest $\chi_\nu^2$.
Based on homology scaling we then assume that the frequencies
in the vicinity of $\CM_{\rm min}^\prime$ can be obtained as
$r \nu_i(\CM_{\rm min}^\prime)$ where $r = [R/R(\CM_{\rm min}^\prime)]^{-3/2}$,
$R$ being the surface radius of a model intermediate between the
actual timesteps in the evolution sequence.
The best-fitting such model is determined by minimizing
\begin{equation}
\chi^2_\nu(r) = {1 \over N-1} \sum_{i=1}^N 
\left( {\nu^{\rm (obs)}_i - r \nu_i(\CM_{\rm min}^\prime) 
\over \sigma_i } \right)^2
\label{eq:chisqfit}
\end{equation}
as a function of $r$.
The resulting value $r_{\rm min}$ of $r$ defines
an estimate $R_{\rm min}$ of the radius of the best-fitting model along the
given sequence. 
In this way we ensure that the scaled model is intermediate
between two successive timesteps
in the evolution sequence for which frequencies have been calculated.
Linear interpolation to $R_{\rm min}$ then defines the final best-fitting
model $\CM_{\rm min}(\CP_k)$
(which obviously in general does not coincide with a
timestep in the evolution sequence)
for the given set of model parameters $\{\CP_k\}$,
and with corresponding $\chi_{\nu,\rm min}^2(\CP_k)$ and 
$\chi_{\rm min}^2(\CP_k)$.
To obtain the final best-fitting model we find the parameter
set corresponding to the smallest 
$\chi_{\nu,\rm min}^2(\CP_k)$ (or $\chi_{\rm min}^2(\CP_k)$)
amongst all the evolution sequences.
The best-fitting frequencies, e.g., for comparison with the observations
are obtained by applying the appropriate scaling
$r_{\rm min}$ to the frequencies of the model $\CM_{\rm min}^\prime$
in the minimizing sequence.

An important goal of the fit is obviously to obtain statistically well
characterized estimates of the stellar properties,
in particular density, radius, mass and age.
In the present analysis these were determined 
as averages and standard deviations
of the properties of the models $\CM_{\min}(\CP_k)$,
over the parameters $\{\CP_k\}$, with the weights
$\chi_{\rm min}^{-2}(\CP_k)$.

The calculations used the latest OPAL equation of state tables
\citep[see][]{Rogers1996} and OPAL opacities at temperatures
above $10^4$\,K \citep{Iglesi1996};
at lower temperature the \citet{Fergus2005} opacities were used.
Nuclear reactions were calculated using the NACRE parameters
\citep{Angulo1999}.
Diffusion and settling of helium was treated using the
\citet{Michau1993} approximations;
diffusion and settling of heavy elements was not taken into account.
Convection was treated using the \citet{Bohm1958} mixing-length formulation.
Although some of the relevant models have a small convective core,
core overshoot was not considered.

The spectroscopically determined ${\rm [Fe/H]} = 0.12 \pm 0.04$
(cf.\ Section \ref{sec:spectParam})
is related to the model quantities $X$ and $Z$ by
\begin{equation}
{\rm [Fe/H]} = \log \left({Z/X \over Z_\odot/X_\odot} \right) \; ,
\end{equation}
where `$\odot$' denotes solar values.
To obtain the composition from this one clearly needs to assume a value
of $Z_\odot/X_\odot$ and a value of $X$ or a relation between $Z$ and $X$.
For the former the \citet{grev93} value of $Z_\odot/X_\odot $ = 0.0245 is used.
We recognize that substantially lower values have been obtained in more
recent solar spectroscopic analyses (see \citet{aspl09} and references
therein).  However, these determinations lead to solar models showing a
substantial increase in the discrepancy with the helioseismically 
determined solar structure (e.g., \citet{bahc05}, \citet{jcd09}), compared
with models based on the \citet{grev93} composition.  Given that the 
reasons for this discrepancy are so far unknown, we prefer to use as
reference a solar composition which provides reasonable agreement
with the helioseismic inferences.
As a model of Galactic chemical evolution
$\Delta Y = 2 \Delta Z$ has often been used,
and hence (fixing the relation roughly to the Sun),
that $X = 0.7679 - 3 Z$.
The grid in composition allows for a spread in \feh \ using models
with ${\rm [Fe/H]} = 0.02$, $0.1$ and $0.18$, consistent with the
\feh \ error of 0.074 after inclusion of systematics.
With the transformation discussed above this corresponds to
$(Z, X) = (0.0183,0.7130), (0.0217, 0.7029)$ and $(0.0256, 0.6910)$.
To avoid being restricted to a specific relation between $X$
and $Z$ we have computed models for all nine resulting
combinations of $X$ and $Z$.

The values of the mixing length have been chosen as
$\alpha_{\rm ML} = 1.5, 1.8$ and $2.1$.
Models were computed initially between $0.9$ and $1.2 \, M_\odot$
with a step of $0.02 \, M_\odot$.
The grid in mass was later refined, with a step of $0.01 \, M_\odot$,
in the vicinity of the best-fitting models.
A total of 282 evolution sequences, with typically 200 -- 300 models in
each, are considered in the fit.

The present use of adiabatic frequency calculations, and an 
inadequate modeling of the near-surface layers, 
cause errors in the resulting frequencies which must be taken
into account in the fit.
Here we use a correction to the frequencies
for these near-surface errors, of the form
\begin{equation}
\delta \nu = a (\nu/\nu_0)^b 
\end{equation}
\citep{Kjelds2008}, where $b = 4.90$ is obtained from a corresponding solar fit,
$\nu_0 =  2071.33 \muHz$ was fixed in the middle of the observed frequency range
and $a = -1.527 \muHz$ was obtained from a fit of a suitable reference model
to the observed frequencies.

With these procedures, the best-fitting model had
$\chi^2 = 5.0$ (cf. Eq. \ref{eq:chisq}).
The quality of the fit is illustrated in the \'echelle diagram \citep{grec83}
shown in Fig.~\ref{fig:asteroech}.
The stellar parameters and standard deviations
obtained from the weighted averages over the evolution
sequences are shown in Table~\ref{tab:SystemParams}.

As a consistency check the global oscillation parameter values
$\Delta\nu$ = 101.51 
$\pm$ 0.09 $\mu$Hz, and $\nu_{\rm max}$ = 2154 $\pm$ 13 $\mu$Hz,
were derived from the best-fitting (peak bagging) frequencies
and amplitudes of the most prominent peaks providing values quite consistent with
estimations from standard, automated detection codes
(\citet{hekk10}, \citet{jcd10}; \citet{vern11}).
The resulting density, mass and radius provided by grid-based
solutions 
(\citet{basu10}; \citet{karo10})
were consistent with our more detailed fit presented here.
Likewise, fits using the Asteroseismic Modeling Portal (see
\citet{metc09} for details) gives values very close to 
those presented here.

Our primary asteroseismic solutions used the formal spectroscopic
errors of 44 K on \teff \ and allowed a spread of $\pm$ 0.08 dex on \feh.
We have used 
grid-based solutions starting with these errors, and the more
appropriate errors of 74 K on \teff \ and 0.074 dex on \feh \
to show that changes
in the directly constrained stellar density are negligible, and that
inferred values of stellar mass and radius change by    
0.2 and 0.3 $\sigma$ respectively in comparison to the errors already quoted in Table 4.
The associated errors on stellar properties also changed little in 
comparison to values already in use.

The asteroseismic solutions given here are based on only 
about half of the now available short cadence data.
However, the quality of the frequency extractions for \starname\
with the data through Q7 only is already sufficiently good
that residual errors from the asteroseismic solution are
negligible for the inference of exoplanet parameters.
Indeed the asteroseismology for \starname\ will be among the
very best possible with \ek, and further results concentrating
on fine details of this star will appear in the future.
As an example, some of the best fitting stellar evolution
models indicated a small convective core, hence it would 
be good to explore inclusion of convective core overshoot.
Initial exploration of core overshoot indicated that this
had negligible impact on the mean stellar density required
for exoplanet inferences and was not further pursued.
The quality of \starname\ data for asteroseismology will likely
support inferences on the outer convection zone depth \citep{mazu13}
and obliquity of the rotation axis \citep{chap13} in future analyses.

\starname\ is a near twin of $\alpha$ Cen A with inferred values of
\teff, \rstar, \mstar, \lstar, and age all within 1-$\sigma$ of each other,
despite the small error bars for both -- see \citet{bazo12} and 
references therein.  Of fundamental parameters only the [Fe/H]
differs significantly with $\alpha$ Cen A being more
metal rich at 0.24 $\pm$ 0.02 \citep{neuf97} compared to \starname\ at 0.12 $\pm$ 0.074.
Both stars have now been interpreted with the benefit of asteroseismology.
The quality of asteroseismic constraints are superior for \starname, with
general astronomical knowledge being better for the nearby binary $\alpha$ Cen A.
Differential analyses of these two interesting stars may prove fruitful.

\section{Planet Characteristics}
\label{sec:planet}

\subsection{Fits to photometry and radial velocities}

The physical and orbital properties of both transit signatures are
derived by simultaneously fitting \ek\ photometry and Keck radial
velocities and by adopting the mean stellar density, \rstar, and \mstar\
of the host star as determined by asteroseismology.

The \ek\ photometry and Keck radial velocities are fit with
non-interacting Keplerian orbits.  The model parameters are the mean
stellar density (\rhostar) and a flux and radial velocity zero point and
for each planet, the time of transit ($T0$), orbital period ($P$), impact
parameter ($b$), scaled planetary radius (\rprs), radial velocity
amplitude ($K$) and eccentricity and argument of
pericenter  parametrized  as \ecosw\ and \esinw. The transit was
modeled  using the analytic formalization of Mandel \& Agol (2002) to fit
photometric observations of the transit. We  use the quadratic
parameterization of limb darkening  also described by Mandel \& Agol (2002)
with  coefficients (0.4096, 0.2602) calculated by Claret \& Bloemen (2011)
for the \ek\ bandpass.   Model fits to the \koib\ light curve yield an
eccentricity that is consistent with zero  (\ecosw\ = 0.02 $\pm$ 0.10;
\esinw\ = 20.13 $\pm$ 0.20) which is  consistent with our expectations
for tidal circularization  (Mazeh 2008). Given the large orbital
separation of the  outer planet candidate, we can not assume its orbit
to  be circular based on tidal circularization.  The duration  of the
transit for the outer planet candidate is consistent  with a circular
orbit, but the resulting upper limit is still  significant ($e$
$\simeq$ 0.2).  For the remainder of our discussion, the models are
constrained to zero eccentricity for  both \koib\ and \koic. The
radial velocity variations  are modeled by assuming non-interacting
(Keplerian)  orbits. With \koic\ validated, the  relative
inclination between the two orbits is likely less  than 20 degrees, as
larger relative inclinations would require a  fortuitous alignment of the
orbital nodes for both planets  to transit (Ragozzine \& Holman 2010).

We initially fit our observations by fixing $\rhostar$ to its
asteroseismic values (see Section 6.2). Model  parameters are found by
chi-squared minimization using  a Levenberg-Marquardt prescription. We
then use  the best-fit values to seed a Markov Chain Monte Carlo  (MCMC)
parameter search (Ford 2005) to fit all model  parameters with \rhostar\
from the asteroseismic solution. We adopt the asteroseismic
determined  mean-stellar density as a prior of the overall solution.

\subsection{Transit Timing Variation prospects}

For a nominal two-planet model (i.e., circular orbits with masses
from Table 4), the predicted root mean square (RMS) transit timing
variations (TTVs) for \koib\ \& \koic\ are both less than half
a minute.
The median timing uncertainties (based on long-cadence observations)
are ~3.3 and ~20 minutes.
Thus, it is not surprising that \ek\ has not provided a TTV signal due
to the interaction of \koib\ and \koic\ from initial searches.
Even with a possible factor of two precision gain from short-cadence
data, and an available long time-series providing square root of the
number of transits gain, predictions are that detection of a TTV
signal on \koib\ remains marginal, and unlikely for \koic.
Even increasing the masses of both planets by three times the upper
``one-sigma'' uncertainty above the estimates in Table 4, the
predicted RMS TTVs are less than a minute for both planets.
Similarly, even models with unrealistically large eccentricities
($e$ = 0.3 and the nominal masses) can result in RMS TTVs of less than a
minute.
Therefore, we have not performed a detailed TTV analysis of this system.

\subsection{Composition of \koib}

The synthesis of radial velocity monitoring, transit photometry,
and precise asteroseismic stellar characterization reveals that
\koib's mass, radius, and density are all intermediate between
the properties of Earth and the Solar System ice giants. 
\koib's bulk density ($3.32^{+0.86}_{-0.98}~\rm{g\,cm^{-3}}$)
is low enough to imply that volatiles (in the form of H/He or
astrophysical ices) make a significant contribution to the total
planet mass and volume. \koib\ cannot be composed of iron and
silicates alone; an iron-poor silicate composition is too dense
by more than $3~\sigma$. Even a carbon-rich mineralogy --
which may lead to solid planets with larger radii than the
Earth-like mineralogy often assumed \citep{madh12} --
does not account for the planet density within $1~\sigma$.
Following \citet{roge10} and \citet{roge11},
we constrain the range of bulk compositions that are consistent
with \koib's measured mass and radius. 
Assuming an Earth-like rocky interior composition (consisting of
32\% Fe and 68\% silicate by mass) \koib\ would need
between 0.07\% and 0.6\% of its mass in a H/He gas layer,
or alternatively between 21\% and 76\% of its mass in a
water vapor envelope. Intermediate compositions with a
mixture of H/He and higher mean molecular weight material from ices are also possible. 

Compared to \koib, the compositions of the other planets
orbiting \starname\ are less well constrained because the bulk planet densities are unknown.
Given its Jupiter-like minimum mass, the outermost planet
(\koid) is likely to be dominated by H/He. Without a transit
measurement of \koid's radius, however, neither the dominant
composition nor the more subtle proportion
of heavy elements in the planet interior can be directly inferred. 
At the other extreme of the planetary mass scale, Earth-sized
\koic\ has only a marginal radial velocity detection. 
Planet interior structure models place more stringent
constraints on the planet mass than the $2~\sigma$
radial velocity upper limit of $10.6~M_\oplus$.
If \koic\ is a rocky body composed of iron and silicate,
its mass would fall within $0.65~M_\oplus < M_p < 2.5~M_\oplus$ --
even a pure iron configuration is allowed.
A residual radial velocity
precision better than $70~\rm{cm\,s^{-1}}$ is needed to constrain \koic's make-up. 

A striking feature of the \starname\ planetary system is that it
harbors one of the most strongly irradiated volatile-rich mini-Neptunes detected to date. 
The stellar energy flux received by \koib\ is more than
$412\pm34$ times larger than that received by the Earth. 
Among low-mass ($M_{\rm P}<20~M_{\oplus}$) planets with measured radii,
only Kepler-10b, 55~Cnc~e, CoRoT-7b, and Kepler-18b are more
strongly irradiated by their host stars. 
These planets have higher densities than \koib, however,
and are consistent with volatile-less compositions within
the $1~\sigma$ uncertainties on their measured masses and radii.
Kepler-10b, CoRoT-7b, and Kepler-18b may be comprised solely of
iron and silicates (with no H/He or astrophysical ices),
and 55~Cnc~e (although not dense enough to have a silicate
composition) could have a carbon-rich solid composition without
a volatile envelope \citep{madh12}.
\koib's status as the most strongly irradiated mini-Neptune
that unambiguously (when the $1~\sigma$ uncertainties are taken
into account) has a significant amount of volatiles makes it a
valuable benchmark for planet mass-loss models. 

Like many of the highly irradiated \ek\ planets,
mass loss has likely had an important influence
sculpting Kepler-68b's composition over its 6.3 Gyr lifetime.
Indeed, \koib\ lies near the edge of the empirical mass
loss destruction threshold noted by several authors
\citep{leca07, ehre11, jack12, lope12}.
At 8.3 $M_{\mathrm{\oplus}}$, \koib\ is close to the
minimum mass of 6.5 $M_{\mathrm{\oplus}}$ predicted by mass
loss in \citet{lope12}. In order to examine the vulnerability
of \koib\ in greater detail we employed the coupled thermal
and mass loss evolution models of \citet{lope12}, assuming
an Earth-like core and 10\% mass loss efficiency. 

Although, \koib\ is stable against mass loss today, it
is possible that it underwent substantial mass loss early
in its history when radii were larger and stellar XUV
fluxes were over 100$\times$ higher \citep{riba05}.
In fact, if \koib\ had a H/He envelope then, it was
vulnerable to a type of runaway mass loss that occurs when
the mass loss timescale becomes short compared to the cooling
timescale \citep{bara04, lope12}. Although less than 1\% H/He
today, \koib\ would need to have been $\sim$80\% H/He
when it was 10 Myr old in order to have a residual primordial
H/He envelope today.  Moreover, models that undergo this
type of extreme mass loss almost always lead to a planet
completely stripped of its H/He envelope \citep{lope12}.
The initial conditions must be carefully fine tuned in
order to arrive at a planet that has such a small but
non-zero H/He envelope today. This suggests that
it is unlikely that \koib\ retains a primordial H/He envelope.

On the other hand, a steam envelope on \koib\ should
be very stable against mass loss. If \koib\ is $\sim$50\%
water today, then it has only lost $\sim$1\% of its initial
water envelope since it was 10 Myr old. This suggests that
it is possible that like Kepler-68d, 68b formed beyond the
snow-line and migrated to its current orbit. However, another
distinct possibility is that \koib\ does in fact have a
H/He envelope, just not a primordial one. \citet{elki08}
showed that rocky planets could outgas up to 0.9\% of their mass
in H/He, more than sufficient to explain the envelope needed today.

On the whole, the \starname\ planetary system shares characteristics
both with the compact Kepler multi-planet systems (e.g., Kepler-11
and Kepler-20) and with the Solar System.
Like Kepler-11 and Kepler-20, \starname\ has multiple transiting planets
within 0.1~AU. In common with the Solar System, \starname\ has a
Jovian-mass planet residing outside (at greater orbital separations than)
the smaller bodies in the inner system. 
Between 0.1 and 1.4 AU, there are no confirmed planets in the
\starname\ system, although our census may be incomplete. 
The presence of a volatile-rich super Earth within 0.06~AU
combined with the presence of a luke-warm Jupiter inside the
snow line makes the \starname\ system an interesting case study
for planet formation and migration theories. 

\section{Summary}
\label{sec:summary}

Two distinct sets of transit events were detected in the lightcurve
of \starname \ constructed from $\sim 2$ years of \ek\ photometry. 
Physical models, constrained by the asteroseismology-derived
stellar parameters, were simultaneously fit to the transit light curves
and the precision Doppler measurements.  Modeling produced tight constraints
on the properties of \planetb: $\mpl=\mplanetb$ \mearth, \rpl=\rplanetb$ \rearth,
\rhopl=\rhoplanetb$ g cm$^{-3}$.  Evaluation of these properties
within a theoretical framework allowed us to draw conclusions
about the planet's composition, arguing that a simple iron and 
silicate structure is excluded.
\koib\ must retain significant volatiles, even though highly irradiated.

The outer planet, \koid, is detected only in radial velocities
for which an upper limit to the mass is approximately Jupiter
in scale, and in a Mars-like orbit.
Transits of \koid\ would not be expected, if at the same inclination
inferred from the impact parameters for \koib, and \koic.
At the two epochs of \ek\ 
data in which transits of \koid\ would be seen if they existed, none are present.
There is still a 1\% chance that if \koid\ does transit, 
the transits could have been missed in minor data gaps.  The presence
of an outer giant planet further enriches the interpretive 
potential for the \starname\ system.

\koic, the intermediate planet, produces transits of only
55 ppm depth (less than 2/3 that of an Earth analog).  But
due to the brightness of \starname, coupled with low stellar
activity and the modest orbital period of 9.6 days, it is detected
at high confidence from the \ek\ photometry.  Radial 
velocities do not provide confirmation, although the formally
inferred amplitude at a phase fixed from the transits appears
at the 1-$\sigma$ level.  The upper limit on mass from the RVs
is not significant for inferring interesting aspects about 
the planet composition.
\blender\ provided a sufficiently high odds ratio to assert
tha \koic\ is a planet.  This validation came with
an added complication in this case, however, in that some individual false
positive scenarios (the best being for a 7 magnitude fainter
background eclipsing binary) provided formally better fits
to the phased light curve than did a simple planet transit model.
With an odds ratio over 10,000 for the planet interpretation,
it is proper to accept this, with perhaps some qualification reserved
in this case.  Were further data to more definitively provide 
evidence that a subtle feature at phase 0.5 in the light curve
is properly interpreted as a secondary eclipse, then the
statistical argument for ``validation" would be over-ruled by
``confirmation" as a false positive.

This qualifies as an interesting system even in the context of
so many exciting and unique discoveries coming from the 
{\em Kepler Mission} -- with a bright, quiet star providing exquisite asteroseismic
constraints on stellar properties; radial velocities providing a precise mass for
one transiting planet and supporting discovery of an outer 
planet; and the second transiting planet validated.
That the innermost transiting planet
has been shown to have a density intermediate between 
terrestrial and gas giant planets, with sufficient fidelity 
to inform theoretical models of its structure, further bolsters
the assertion that the \starname\ exoplanet system is an 
important development in this rapidly expanding field.

\acknowledgements 

Funding for this tenth Discovery mission is provided by NASA's Science Mission Directorate.
The many people contributing to the development of this mission
are gratefully acknowledged.
We thank Elizabeth Adams, Eric Agol, Natalie Batalha, William Borucki,
Stephen Bryson, William Cochran, Andrea Dupree, Debra Fischer, 
Christopher Henze and David Monet for discussion and contributions.
The anonymous referee made comments serving to improve the paper.
Some of the data used here were
obtained at the W.M. Keck Observatory, which is operated as a
scientific partnership among the California Institute of Technology, 
the University of California, and NASA.
The W.M. Keck Foundation provided generous financial support 
to the Keck Observatory.  This work is also based in part on
observations made with the {\em Spitzer Space Telescope} which
is operated by the Jet Propulsion Laboratory, California Institute 
of Technology under a contract with NASA.
Partial support for this work was provided by NASA through an award issued by JPL/Caltech.
Support for L.A.R. was provided through Hubble Fellowship grant \#HF-51313.01-A
awarded by the Space Telescope Science Institute, which is operated by the
Association of Universities for Research in Astronomy, Inc., for NASA 
under contract NAS 5-26555.
G.T. acknowledges partial support for this work from NASA grant
NNX12AC75G (Kepler Participating Scientist Program).
Funding for the Stellar Astrophysics Centre is provided by 
The Danish National Research Foundation.  The research is 
supported by the ASTERISK project (ASTERoseismic Investigations
with SONG and \ek) funded by the European Research Council
(Grant agreement no.:  267864).
Asteroseismic analysis was supported in part by White Dwarf Research
Corporation through the Pale Blue Dot project (http://whitedwarf.org/palebluedot).
S.B. acknowledges support from NSF grant AST-1105930.
R.L.G. has been partially supported by NASA co-operative agreement: NNX09AG09A.

{\it Facilities:} \facility{{\em Kepler}}.

\clearpage

%TABLES

\begin{table}
\begin{center}
\tablenum{1}
\caption{Direct and Difference Image values for \koib\ and \koic
\label{tab:difftab}}
\begin{tabular}{|c|cccc|cccc|cccc|}
\tableline\tableline
Row & \multicolumn{4}{c|}{Direct} & \multicolumn{4}{c|}{Difference \koib } & \multicolumn{4}{c|}{Difference \koic} \\ \hline
Column & 943 & 944 & 945 & 946 & 943 & 944 & 945 & 946 & 943 & 944 & 945 & 946 \\ \hline
343 & .000 & {\em .003} & {\em .001} & .000 & +.000 & {\em +.160} & {\em +.000} & +.000 & -.004 & {\em -.026} & {\em +.000} & +.001 \\
342 & .003 & .111 & .099 & .002 & +.003 & -.004 & +.163 & +.001 & -.013 & +.000 & +.286 & +.008 \\
341 & .005 & .120 & .111 & .015 & +.006 & +.048 & -.002 & +.011 & -.005 & -.078 & -.005 & +.035 \\
340 & .009 & .124 & .124 & .017 & +.008 & +.008 & +.010 & +.017 & -.002 & -.013 & +.020 & +.037 \\
339 & .003 & .111 & .124 & .010 & +.002 & +.257 & +.217 & +.008 & +.002 & +.169 & +.385 & +.021 \\
338 & .000 & .001 & .003 & .003 & +.000 & +.002 & +.009 & +.003 & +.004 & +.002 & +.018 & -.005 \\
\tableline
\end{tabular}
\tablecomments{Column and row values refer to pixels on channel 59
of the \ek\ detectors for Quarter 9.
The Direct image has been normalized by the total electrons per
cadence of 2.4$\times 10^9$.
The difference images have been normalized by the same factor scaled
by the known depths of 346 and 53 ppm for \koib\ and \koic\ respectively.
For internal consistency the sums over these normalized difference
images should be $\sim$1 (this is satisfied).
\starname\ is nearly centered in these tabular domains, while KIC 11295432
(7 magnitudes fainter companion -- see text) is near center of row 343, and 
columns 944--945 pixels as italicized in entries.
Typical uncentainties for \koib\ entries are 0.004, and 0.04 for \koic.}
\end{center}
\end{table}
\clearpage

\begin{table}
\begin{center}
\tablenum{2}
\caption{Relative Radial Velocities and Line Bisectors for \starname
\label{tab:radtab}}
\begin{tabular}{crrrr}
\tableline\tableline
HJD & RV & RV$_e$ & bs & bs$_e$ \\
-2450000 & (\ms) & (\ms) & (\ms) & (\ms) \\
\tableline
 5313.082 &    3.98 &    2.1 &   6.1 & 1.8 \\
 5319.109 &    0.00 &    2.3 & -10.7 & 2.7 \\
 5322.051 &   -2.75 &    1.9 &  -2.1 & 1.8 \\
 5372.983 &   -1.27 &    1.9 &  17.3 & 2.5 \\
 5377.929 &   -0.39 &    1.9 &  -2.5 & 4.4 \\
 5381.000 &   -7.85 &    2.0 &   4.5 & 2.8 \\
 5396.963 &   -5.20 &    2.1 &  -7.7 & 1.4 \\
 5400.020 &    0.14 &    3.4 &  -3.1 & 2.7 \\
 5412.923 &   -8.94 &    1.9 &   9.6 & 3.9 \\
 5426.913 &   -6.81 &    1.9 &  -3.9 & 3.7 \\
 5431.784 &   -1.04 &    1.9 &  -8.4 & 3.3 \\
 5434.873 &  -14.06 &    1.7 &   7.0 & 3.5 \\
 5435.931 &   -9.75 &    2.0 &   2.3 & 3.5 \\
 5436.971 &   -8.34 &    1.9 &  -3.9 & 3.0 \\
 5437.945 &   -4.90 &    1.8 &   3.1 & 2.3 \\
 5438.996 &   -6.18 &    1.8 &   4.8 & 2.0 \\
 5439.928 &  -10.42 &    1.7 &   9.8 & 1.8 \\
 5440.975 &  -10.20 &    1.7 &  -7.6 & 1.2 \\
 5455.810 &  -12.90 &    2.0 &  -3.8 & 1.5 \\
 5490.830 &   -8.52 &    2.0 &  -2.0 & 3.0 \\
 5672.026 &   27.25 &    1.9 &  -3.5 & 3.5 \\
 5672.998 &   23.74 &    2.0 &   1.6 & 3.1 \\
 5673.996 &   28.12 &    2.0 &   6.5 & 2.1 \\
 5696.974 &   30.91 &    2.0 &  11.1 & 2.2 \\
 5697.964 &   33.04 &    2.1 &   9.2 & 1.8 \\
 5698.962 &   26.37 &    2.0 &  13.0 & 1.2 \\
 5722.995 &   31.84 &    2.0 &  -5.9 & 3.0 \\
 5724.034 &   36.50 &    2.1 & -10.0 & 2.4 \\
 5728.901 &   32.98 &    2.1 &   3.2 & 2.2 \\
 5734.064 &   34.50 &    2.0 &  -5.3 & 2.2 \\
 5734.951 &   35.37 &    2.0 &  -3.7 & 2.7 \\
 5735.975 &   33.84 &    2.0 &   2.7 & 1.3 \\
 5739.034 &   33.61 &    2.0 &   0.8 & 1.8 \\
 5751.797 &   27.85 &    2.1 & -14.9 & 2.5 \\
 5752.105 &   27.06 &    2.0 & -10.2 & 2.5 \\
 5752.779 &   25.05 &    2.0 &  -2.8 & 1.4 \\
 5759.975 &   27.91 &    2.0 &  15.1 & 1.7 \\
 5761.076 &   30.96 &    1.9 &  -6.6 & 2.9 \\
 5761.842 &   24.79 &    2.0 &  -0.7 & 1.8 \\
 5763.033 &   28.05 &    2.0 &  -9.7 & 1.4 \\
 5763.851 &   24.13 &    2.0 & -10.1 & 2.6 \\
 5782.908 &   24.68 &    2.0 &   2.5 & 2.0 \\
 5795.024 &   16.72 &    2.2 &  -8.7 & 1.6 \\
 5814.736 &   22.27 &    1.9 &   9.0 & 2.7 \\
 6077.045 &   -6.84 &    2.0 & -10.0 & 1.7 \\
 6098.094 &   -2.72 &    2.1 &  -4.9 & 1.5 \\
 6098.829 &   -3.40 &    2.0 &  -5.2 & 1.8 \\
 6102.008 &    3.17 &    2.0 &  18.4 & 1.9 \\
 6114.872 &   -6.21 &    2.0 &  -0.4 & 2.3 \\
 6145.875 &   -0.55 &    2.1 &  10.2 & 2.0 \\
 6148.929 &    5.57 &    1.9 &   1.8 & 1.3 \\
 6151.061 &   -1.64 &    2.0 &  -1.7 & 1.5 \\
\tableline
\end{tabular}
\end{center}
\end{table}
\clearpage

\begin{table}
\begin{center}
\tablenum{3}
\caption{Measured Frequencies $\nu_{nl}$ of \starname\ (in $\mu$Hz).
\label{tab:freqs}}
\begin{tabular}{cccc}
\tableline\tableline
$n$ & $l$ = 0 & $l$ = 1 & $l$ = 2 \\
\tableline
14 & -- & 1661.02 $\pm$ 0.35 & -- \\
15 & 1668.43 $\pm$ 0.29 & 1713.38 $\pm$ 0.15 & 1761.36 $\pm$ 0.32 \\
16 & 1767.09 $\pm$ 0.29 & 1813.49 $\pm$ 0.19 & 1861.58 $\pm$ 0.37 \\
17 & 1867.94 $\pm$ 0.17 & 1914.65 $\pm$ 0.20 & 1962.99 $\pm$ 0.20 \\
18 & 1969.08 $\pm$ 0.14 & 2016.27 $\pm$ 0.11 & 2064.85 $\pm$ 0.11 \\
19 & 2070.73 $\pm$ 0.09 & 2117.70 $\pm$ 0.07 & 2166.32 $\pm$ 0.23 \\
20 & 2172.02 $\pm$ 0.13 & 2219.40 $\pm$ 0.14 & 2268.08 $\pm$ 0.20 \\
21 & 2273.37 $\pm$ 0.15 & 2321.09 $\pm$ 0.14 & 2369.55 $\pm$ 0.60 \\
22 & 2375.43 $\pm$ 0.25 & 2423.57 $\pm$ 0.29 & 2472.54 $\pm$ 0.64 \\
23 & 2477.86 $\pm$ 0.36 & 2525.73 $\pm$ 0.30 & -- \\
\tableline
\end{tabular}
\end{center}
\end{table}
\clearpage

\begin{table}
\begin{center}
\tablenum{4}
\caption{Star and planet parameters for the \starname\ system.
\label{tab:SystemParams}}
\begin{tabular}{lcc}
\tableline\tableline
Parameter & Value & Notes \\
\tableline
%\sidehead{\em Transit and orbital parameters: \planetb}
{\em Transit and orbital parameters: \planetb} & & \\
Orbital period $P$ (days)   & \periodb  & A \\
Midtransit time $E$ (BJD)   & \epochb  & A \\
Scaled semimajor axis $a/\rstar$  & \scaledSemiMajb & A \\
Scaled planet radius \rpl/\rstar  & \scaledPlanetRadiusb & A \\
Impact parameter $b$ & \impactb & A  \\
Orbital inclination $i$ (deg) & \inclinationb  & A \\
Orbital semi-amplitude $K$ (\ms) & \semiAmpb  & B \\
Orbital eccentricity $e$   & \eccb   & B \\
Center-of-mass velocity $\gamma$ (\kms)  & \gammaVelb  & B \\

%\sidehead{\em Transit and orbital parameters: \koic}
{\em Transit and orbital parameters: \koic} & & \\

Orbital period $P$ (days) & \periodc & A     \\
Midtransit time $E$ (HJD) & \epochc & A     \\
Scaled semimajor axis $a/\rstar$ & \scaledSemiMajc & A     \\
Scaled planet radius \rpl/\rstar & \scaledPlanetRadiusc & A     \\
Impact parameter $b$ & \impactc & A     \\
Orbital inclination $i$ (deg) & \inclinationc & A     \\
Orbital semi-amplitude $K$ (\ms) & \semiAmpc  & B \\

%\sidehead{\em Observed stellar parameters}
{\em Observed stellar parameters} & & \\
Effective temperature \teff (K)   & \teffsys  & C  \\
Spectroscopic gravity \logg (cgs)  & \loggSME  & C \\
Metallicity \feh  & \fehsys  & C \\
Projected rotation \vsini (\kms)  & \vsiniSME  & C \\
%\sidehead{\em Fundamental Stellar Properties}
{\em Fundamental Stellar Properties} & & \\
Density \rhostar\ (\gcmc) & \rhostarKASC & D     \\
Mass \mstar (\msun) & \mstarKASC  & D \\
Radius \rstar (\rsun) & \rstarKASC  & D \\
Surface gravity \loggstar\ (cgs)  & \loggKASC  & D \\
Luminosity \lstar\ (\lsun) & \lumKASC  & D \\
Absolute V magnitude $M_V$ (mag)  & \absMag     & D \\
Age (Gyr)     & \ageKASC  & D \\
Distance (pc)     & \distance  & D \\ 

%\sidehead{\em Planetary parameters: \planetb}
{\em Planetary parameters: \planetb} & & \\
Mass \mpl\ (\mearth)    & \mplanetb  & A,B,C,D,F \\
Radius \rpl\ (\rearth)    & \rplanetb  & A,B,C,D \\
Density \rhopl\ (\gcmc)    & \rhoplanetb  & A,B,C,D,F \\
Surface gravity \loggpl\ (cgs)   & \loggplanetb  & A,B,C,D,F \\
Orbital semimajor axis $a$ (AU)   & \semiMajb  & E  \\
Equilibrium temperature \teq\ (K)  & \teqb   & G  \\

%\sidehead{\em Parameters for candidate: \koic}
{\em Planetary parameters: \koic} & & \\
Mass \mpl\ (\mearth)  & \mplanetc  & A,B,C,D,F       \\
Radius \rpl\ (\rearth) & \rplanetc  & A,B,C,D     \\
Density \rhopl\ (\gcmc) & \rhoplanetc & A,B,C,D,F       \\
Orbital semimajor axis $a$ (AU) & \semiMajc  & E       \\
%Equilibrium temperature \teq\ (K) & \teqc  & G       \\

%\sidehead{\em Planetary parameters: \planetd}
{\em Planetary parameters: \planetd} & & \\
Orbital period $P$ (days) & \periodd  & F        \\
Minimum mass $M_P \, {\rm sin}i \, (M_J)$ & \mplanetd & D,F      \\
Orbital semi-amplitude $K$ (\ms) & \semiAmpd & D,F      \\
Orbital semimajor axis $a$ (AU)  & \semiMajd & D,F      \\
Orbital eccentricity $e$ & \eccd & F        \\
%Equilibrium temperature \teq\ (K) & \teqd & F        \\
\tableline
\end{tabular}
\tablecomments{
A: Based primarily on an analysis of the photometry,\\
B: Based on a joint analysis of the photometry and radial velocities,\\
C: Based on an analysis by D. Fischer of the Keck/HIRES template
spectrum using SME \citep{vale96},\\
D: Based on asteroseismology analysis, \\
E: Based on Newton's revised version of Kepler's Third Law and the results from D,\\
F: Based on radial velocities, \\
G: Calculated assuming a random distribution of Bond albedo over 0.0 to 0.5,
and a random set ranging from zero to full redistribution of heat from day to night sides.}
\end{center}
\end{table}
\clearpage

%%%%%%%%%%%%%%%%%%%%%%%%%%%%%%%%%%%%%%%%%%%%%%%%%%%%%%
%FIGURES

\clearpage

\clearpage

\clearpage

\clearpage

\clearpage

\clearpage

\clearpage

\begin{figure}
\begin{center}
\includegraphics[height=175mm]{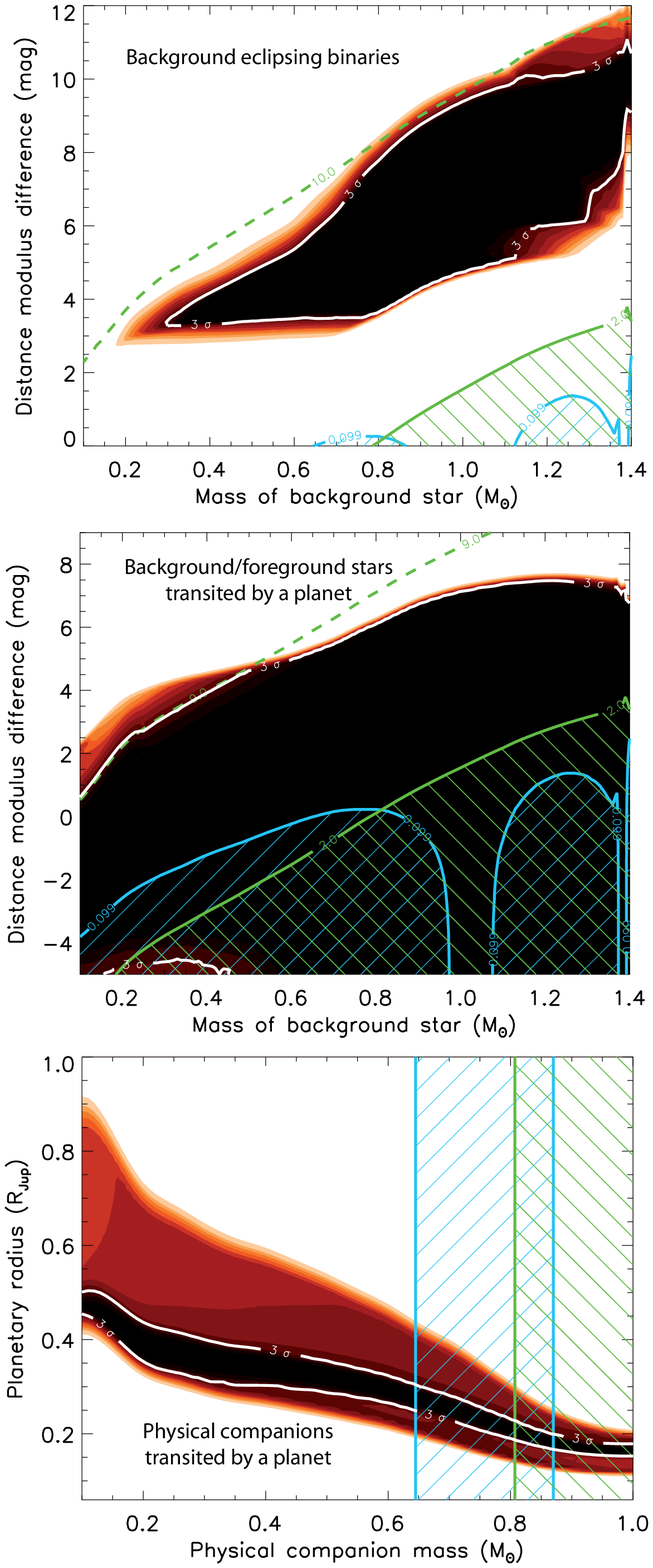}
\end{center}
\end{figure}

\begin{figure}
\begin{center}
\vskip 1pt
\caption{\blender\ goodness-of-fit contours for \koic\ 
corresponding to the three different scenarios that contribute to the
overall blend frequency: background eclipsing binaries (top),
background or foreground stars transited by a planet (middle), and
physical companions transited by a planet (bottom). Only blends inside
the solid white contours produce light curves matching the \ek\
photometry within acceptable limits \citep[3$\sigma$, where $\sigma$
is the significance level of the $\chi^2$ difference compared to a
transit model fit; see][]{Fressin:11}. Lighter-colored areas (red,
orange, yellow) mark regions of parameter space giving increasingly
worse fits to the data (4$\sigma$, 5$\sigma$, etc.), and correspond to
blends we consider to be ruled out. The axes in each panel represent
two of the dimensions of parameter space for blends.  For the top two
diagrams the vertical axis represents the distance between the
background/foreground star and the target, expressed here in terms of
a difference in distance modulus rather than in parsecs.  The
horizontal axis corresponds to the mass (spectral type) of the
intruding star. In the lower panel (physically bound scenarios) the
vertical axis is the size in Jupiter radii of the planet transiting
the companion star.  The cyan cross-hatched areas indicate regions of
parameter space ruled out because the resulting $r-K_s$ color of the
blend is either too red (left) or too blue (right) compared to the
measured color, by more than 3$\sigma$ (0.10 mag). In the top and
middle panels the solid green line is a line of constant magnitude
difference ($\Delta K\!p = 2$) between the target and the background
star. Blends involving stars brighter than this (which lie lower in
the diagram) would have been detected in our spectroscopic
observations (see Sect. 6.1), and are thus ruled out. This is
indicated by the cross-hatched regions below the green lines. Finally,
the dashed green lines in the top two panels are roughly parallel to
the solid green lines, and are also lines of constant magnitude
difference between the target and a background star. They correspond
to the faintest blends that can mimic the transit: approximately
$\Delta K\!p = 10$ for background eclipsing binaries (top), and
$\Delta K\!p = 9$ for background/foreground stars transited by a
planet (middle).\label{fig:blender246.02}}
\end{center}
\end{figure}

\clearpage

\begin{figure}
\begin{center}
\includegraphics[width=150mm]{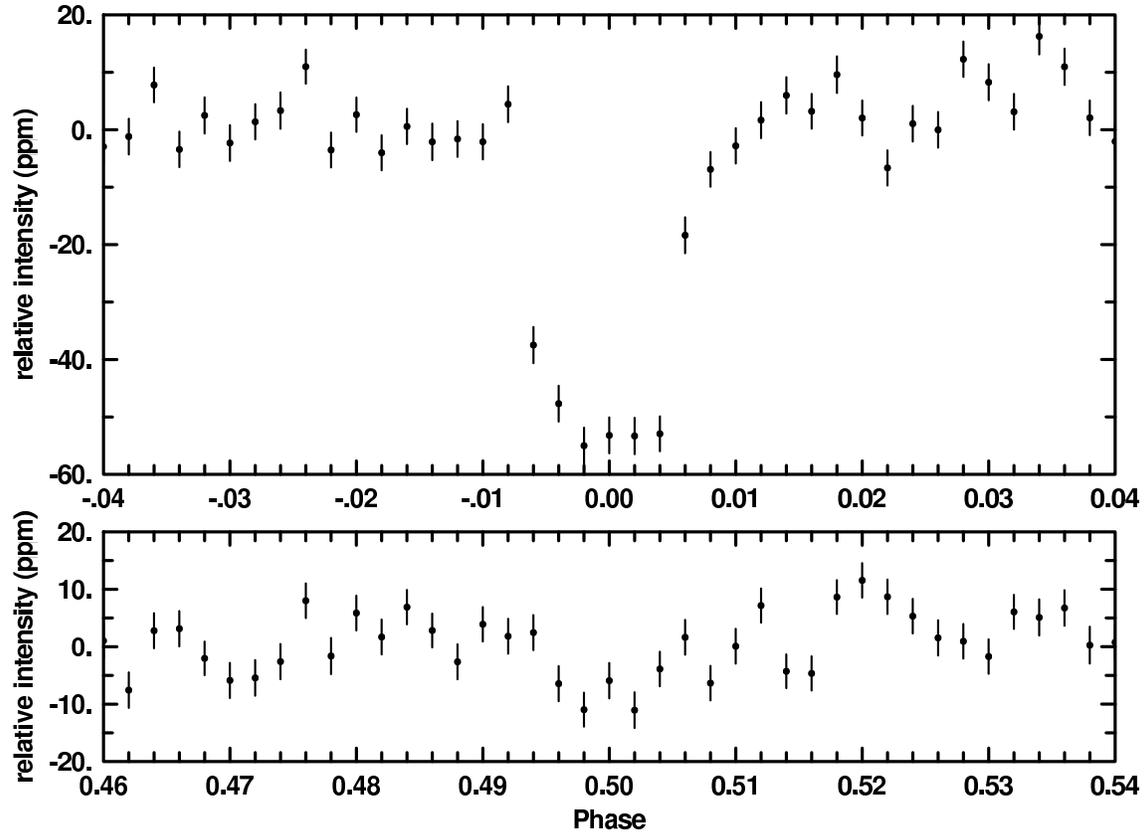}
\end{center}
\caption{The short cadence data for \koic\ phased onto the orbital period.
The upper panel shows the transit centered at phase zero.  The error bars
show the formal error per 0.002 phase bin evaluated as standard deviation
divided by square root of the number of contributing points.  The lower 
panel shows the phased data exactly 0.5 out of phase from the transit and
illustrates the subtle evidence of a secondary eclipse that the \blender\ 
analysis locks onto in providing some false positive scenarios with a 
formally higher significance than a simple transit fit.}
\label{fig:sececlipse}
\end{figure}

\clearpage

\begin{figure}
\begin{center}
\includegraphics[width=150mm]{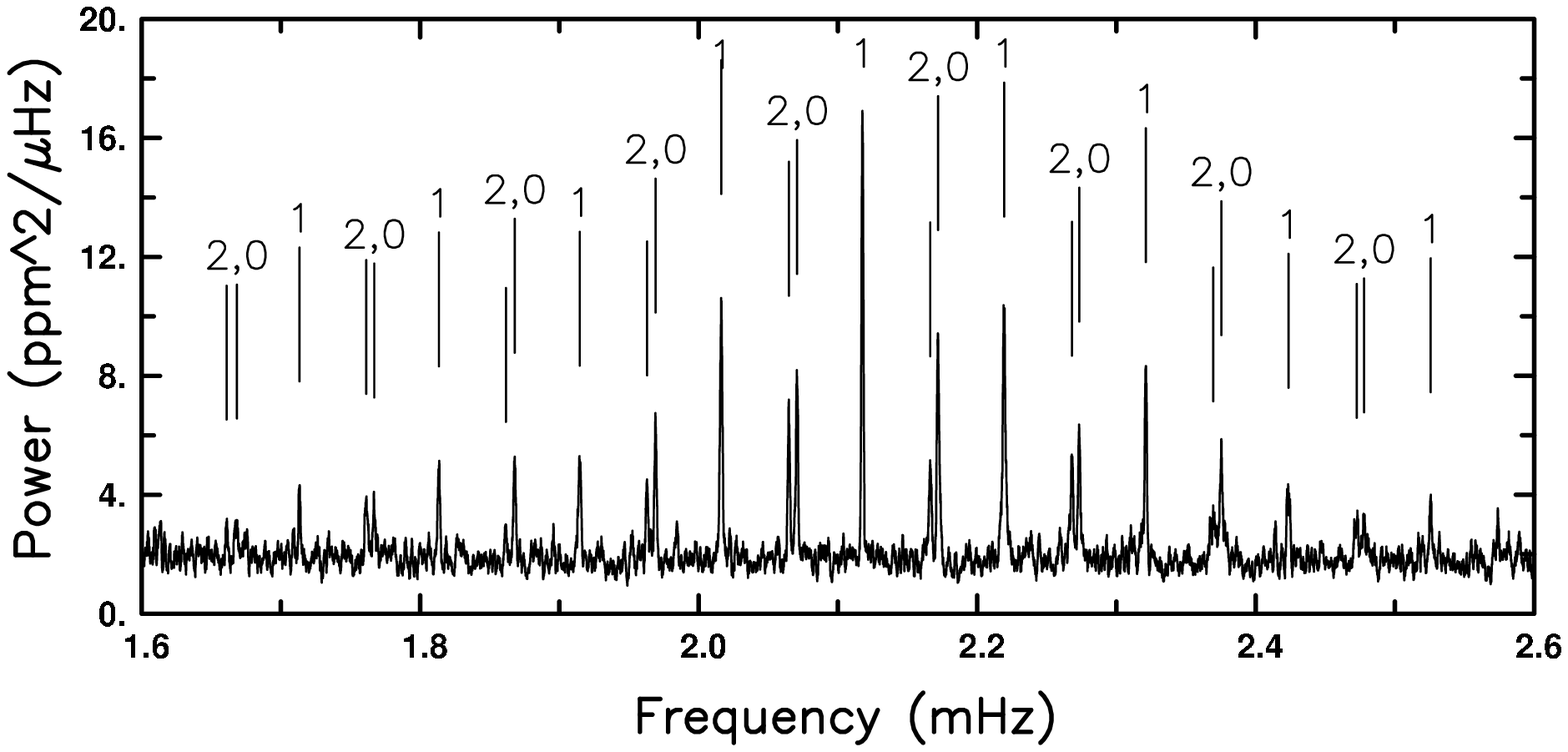}
\end{center}
\caption{Power spectrum for \starname\ showing strong
solar-like p mode oscillations.  Numbers above modes
indicate the angular degree $l$ of each mode used in 
modeling the stellar parameters.}
\label{fig:power}
\end{figure}

\clearpage

\begin{figure}
\begin{center}
\includegraphics[width=150mm]{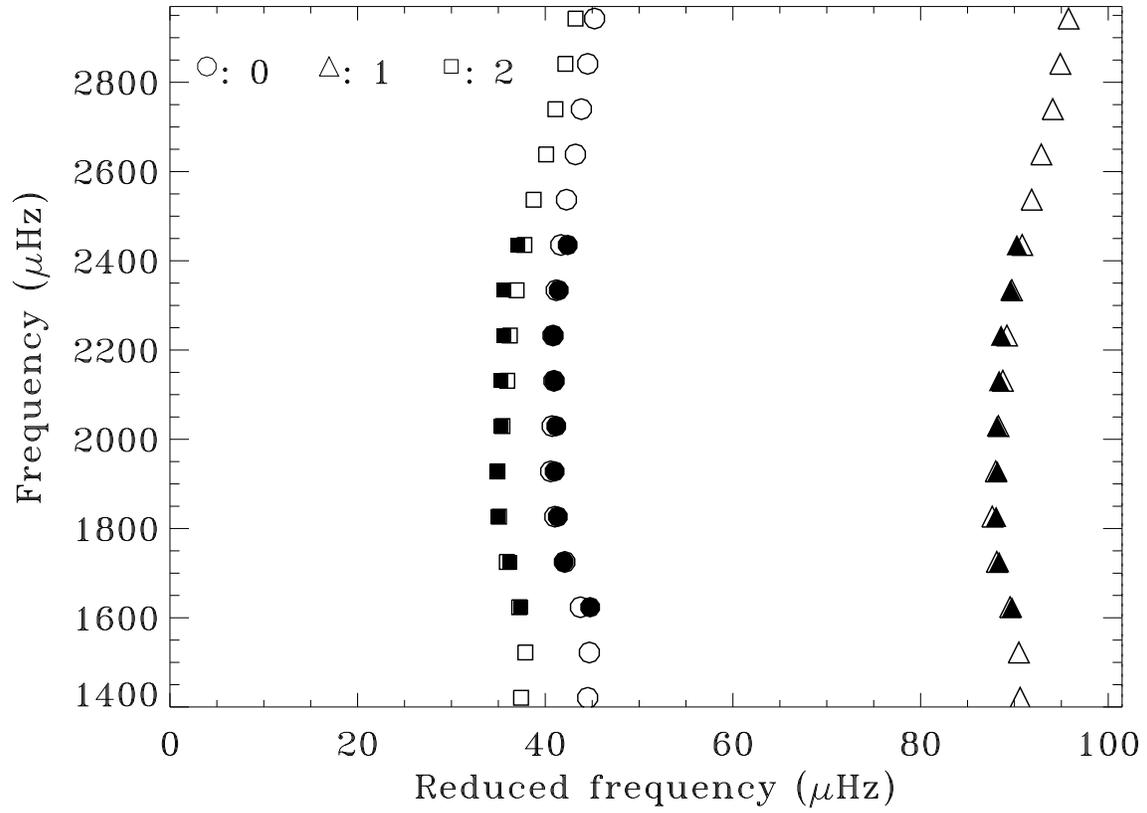}
\end{center}
\caption{Comparison of the observed mode frequencies shown
as solid symbols, and those from the best fitting model
as discussed in Section 6 shown with the open symbols.
The x-axis shows the frequencies after folding by the 
large separation value of 101.51 $\mu$Hz.  Error bars
on the observations are given in Table 3, and are always
smaller than the plot symbols.}
\label{fig:asteroech}
\end{figure}

\end{document}